\documentclass[12pt,preprint]{aastex}
\begin{document}
\title{An Observational Limit on the Dwarf Galaxy Population of the Local Group}
\author{Alan B. Whiting\footnote{Now at the University of Birmingham,
Birmingham B15 2TT, UK.}}
\affil{Cerro Tololo Inter-American Observatory}
\affil{Casilla 603, La Serena, Chile}
\email{abw@star.sr.bham.ac.uk}
\author{George K. T. Hau}
\affil{University of Durham}
\author{Mike Irwin}
\affil{University of Cambridge}
\author{Miguel Verdugo}
\affil{Institut f\"ur Astrophysik G\"ottingen, Germany}

\begin{abstract}
We present the results of an all-sky, deep optical survey for faint Local Group
dwarf galaxies.  Candidate objects were selected from the second Palomar survey 
(POSS-II) and ESO/SRC survey plates and follow-up observations performed to
determine whether they were indeed overlooked members of the Local Group.
Only two galaxies (Antlia and Cetus) were discovered this way out of 206 
candidates.  Based on internal and external comparisons, we estimate that our
visual survey is more than 77\% complete
for objects larger than one arc minute in size and with a surface brightness
greater than an extremely faint limit 
over the 72\% of the sky not obstructed by the Milky Way.  Our limit
of sensitivity cannot be calculated exactly, but is certainly fainter
than 25 magnitudes per square arc second in $R$, probably 25.5 and
possibly approaching 26.
We conclude that there are at most one or two Local Group dwarf galaxies
fitting our observational criteria still undiscovered in the clear part of the
sky, and roughly a dozen hidden behind the Milky Way.  Our work places the
``missing satellite problem'' on a firm quantitative observational basis.
We present detailed data on all
our candidates, including surface brightness measurements.
\end{abstract}

\section{Introduction: Dwarf Hunting}

We present here the final results of our survey for Local Group dwarf galaxies.
These faint objects are heavily selected against in most surveys (for
obvious reasons), but are of disproportionate interest in many areas.
Their external kinematics provides clues to the origin and evolution of the 
Local Group, and internally they show signs of being dominated by dark matter.
Star formation proceeds in different ways among the various known examples,
giving insight into the processes involved.  The total census and luminosity
function provide important contraints on cosmology and structure formation.
For all these reasons, it is worthwhile both to find more examples, and to
set quantitative limits on the completeness and sensitivity of any survey.

The total number of known Local Group dwarfs prior to our survey was
on the order of two dozen, which translates to a very small number per square
degree on the sky.  Unless there were some great undiscovered population
we were faced with the task of searching very large areas in order to have any
reasonable chance of finding more.  And since we are within the Group, new
members could appear in any direction.  These considerations led to an
all-sky effort.  In turn, this meant we had to use photographic survey material,
since no deep all-sky CCD survey yet exists.  

Next came the task of extracting candidate objects from the plate material.
The surveys have been digitized, so one could in principle devise an algorithm 
to pick out likely-looking objects; and this has been done by some groups.
However, at the very low signal levels we were interested in a great number
of false detections would be found, each of which would have to be examined by 
eye anyway; 
so we settled on a visual examination of the survey plates in the first
place.  It quickly became apparent that plate copies of the POSS-II and ESO/SRC 
surveys were superior to film copies in freedom from processing defects which
could mimic dwarf galaxies, so we confined ourselves exclusively to glass.
(This did not turn out to mean a fainter limit on objects found; see below.)
The last consideration involved how to deal with the Milky Way.  A simple cutoff
at a given Galactic latitude would be rather crude and might miss some
dwarfs.  On the other hand, low surface brightness Galactic nebulosity extends
very high in places (we report knots of it at $b \sim45\arcdeg$ below).  In the end 
we decided to examine the entire sky, since even in very extincted regions we
might find (for example) new planetary nebulae which would be of interest to
someone.

The technique as implemented was to search all the 894 north and 894
south survey fields visually, for objects resembling the known Andromeda dwarf 
spheroidals and Tucana, that is of very low surface brightness and large
(one to several minutes of arc in size).  In addition, any object which appeared
to be resolving into faint stars was included.  To screen out plate defects and
reflection nebulae, candidates were required to be on both red and blue copies
of the field.  At the same time, each field was rated as ``good'' (no apparent
Galactic interference), ``troublesome'' (Galactic nebulosity present, but not
covering the whole plate, averaging to something like 50\%), 
or ``poor'' (little or no freedom from the Milky Way)
to provide a rough estimate of Galactic obscuration.  
A total of 338 real objects
were listed from the plate examination.
Catalogs were consulted for these objects,
and those which were known to be Local Group dwarfs, or not, set aside
as non-candidates (we provide a list of those below). 

Follow-up observations were made of these objects using the 1.5m at Cerro Tololo
Inter-American Observatory, the 2.1m at Kitt Peak National Observatory, and
the 2.5m Isaac Newton Telescope at Observatorio del Roque de los Muchachos on
the island of La Palma.
Each was imaged in $R$ in sufficient depth to show the tip of the Red Giant Branch at
a nominal distace of 1 Mpc.  Those objects which resolved into stars were also
imaged in $V$ and $I$ to provide a color-magnitude diagram and thus a determination
of distance.  Time sometimes allowed images also in narrowband H$\alpha$, to
distinguish HII regions from stars and confirm the identification of emission
nebulae.

\section{Results}

In all we had 206 candidates (134 north, 72 south) and 132 non-candidates (92 north,
40 south).  Six objects in the north and six in the south could not be detected
in follow-up observations; this fact will be discussed below.

There is clearly a strong asymmetry, amounting to a factor of two between the north and the
south.  This cannot not be traced, as in much of astronomy, 
to a matter of historical development or to differences in equipment or
techniques.  We believe it to happen because a large number of Local Group 
dwarf candidates must actually be larger,
more distant galaxies, but not {\em much} more distant; they will be in 
nearby groups.
There are simply more of these, and more populous ones, in the 
northern sky, plus of course the Virgo Cluster.

Observational data on Local Group dwarf galaxy candidates from the 
POSS-II survey are
presented in Table \ref{northcand} and from the ESO/SRC plates in 
Table \ref{southcand}.
All data were derived from our follow-up observations.  Table \ref{southcand} is
an updated version of that published in \citet{WHI02}, and where different
supercedes that listing.
Objects selected on the plates but rejected as candidates
for various reasons are listed in Table \ref{northnon} for the north
and Table \ref{southnon} for the south.

The first column in each table is the adopted designation\footnote{The field of dwarf
galaxy nomenclature is a mess.  UKS 3232-23, for instance, is a dwarf irregular galaxy 
in the Sculptor Group, but is not the Sculptor Dwarf Irregular Galaxy (sometimes known
as SDIG) nor the Sculptor dwarf spheroidal (sometimes known as just Sculptor); in
turn, SDIG must not be confused with SagDIG, the Sagittarius Dwarf Irregular Galaxy,
and the latter is distinct from Sagittarius (which is much closer and
in the process of disintegration under the action of the Milky Way).  One recently
discovered satellite of M31, the Andromeda galaxy, was named Andromeda VI, continuing
an established series; but since it
is actually in the constellation Pegasus, not Andromeda, 
the alternate name of Pegasus Dwarf Spheroidal came into
use, which however invites confusion with the previous Pegasus (which now must be
distinguished by the added ``Dwarf Irregular'').  The situation is not made any
easier by the fact that many of the objects have several names in different catalogs, and
different groups of workers tend to use different designations (and the fact that the
two major online databases, Simbad and NED, use slightly different formats for the same
catalogs).}.
We have used what seems to be the most popular designations (though other choices could
certainly be made).  Where no previous designation could be found, we have devised one
according to the IAU guidelines. We have not used designations based on non-visual catalogues
(for instance, IRAS point sources) because the identity is not always certain.  
In cases of confusion, the positions provide an
unambiguous guide.

Positions are given for all objects in the 2000 equinox, and are accurate to 0.1 minute
{\em when the object morphology permits}.  Such an accuracy is clearly meaningless for
an amorphous patch of Galactic nebulosity several minutes wide, with no clear center 
or concentration.  Even for more limited and definable objects, like spheroidal galaxies,
the low surface brightness and diffuse nature of our targets means that the centers are
not always clearly defined to within 0.1 minute.  In all cases, however, the position
given falls well within the object and is quite adequate to center it in any subsequent
field of view.

Sizes are those of the object as seen in the follow-up image.  They are not rigorously determined
(for instance, size within a certain surface brightness limit) but are included to provide a 
general indication of size.

Surface brightness was determined in the $R$ band according to a procedure which will be
detailed below.

The identification of the type of object is based on the morphology as seen in the follow-up
$R$ image, sometimes with additional data from other bands or published sources.  While
most are fairly definite, some are certainly arguable and we do not pretend that our listing
is infallible in all cases.

\begin{deluxetable}{lccccl}
\rotate
\tablewidth{0pt}
\tablecaption{Northern Sky Local Group Dwarf Galaxy Candidates \label{northcand}}
\tablehead{\colhead{Name} & \colhead{RA} & \colhead{Dec} & \colhead{Size}
& \colhead{Surface Brightness} & \colhead{Type} \\
& \colhead{(2000.0)} & \colhead{(2000.0)} & \colhead{arc min} & \colhead{$R$ mag arc sec$^{-2}$} & }
\startdata
WHI J0004+52&	00 04 21.1&	+52 37 00.4&	5.7 x 3.2&	$23.6 \pm 0.5$		&Gal. neb.\\ 
WHI J0018+65&	00 18 05.6&	+65 50 01.0&	2.5 x 2.7&	$23.5 \pm 0.2$	&	PN\\
WHI J0039+66 &   00 39 26.3 &     +66 51 55.6&     3.6 x 3.2&       $24.0 \pm 0.05$&		emission neb.\\
WHI J0040+66  &  00 40 27.1&      +66 55 32.2 &    4.8 x 6?&        $24.0 \pm 0.4$ & 	Gal. neb.\\
WHI J0144+17&	01 44 33.3&	+17 43 00.1&	2.0 x 1.0&	$25.5 \pm 0.3$	&	Gal. neb.\\
WHI J0156+63&	01 56 11.1&	+63 48 15.4&	6.5 x 3.7&	$24.2 \pm 0.3$	&	Gal. neb.\\
WHI J0156+13&	01 56 46.4&	+13 37 26.8&	2.6 x 1.9&	$24.2 \pm 0.2$	&	Gal. neb.\\
ZOAG G133.63-03.62	&02 12 17.1&	+57 33 59.3&	0.7	&	$23.0 \pm 0.4$&		spiral\\
WHI J0226+52&	02 26 41.5&	+52 36 57.6&	$>8.6$x9.3&	$24.1 \pm 0.3$	&	Gal. neb.\\
KK98-20	&	02 34 30.3&	+22 34 43.3&	0.9 x 0.8&	$24.5 \pm 0.6$	&	faint spiral\\
ZOAG G134.31+06.24&      02 49 11.5&      +66 28 39.7&     1.6 x 1.0&       $22.3 \pm 0.05$&	face-on spiral\\
ZOAG 139.32+04.85&	03 23 05.0&	+62 47 09.8&	1.6 x 0.7&	$24.4 \pm 0.2$	&	emission neb.\\
KKH01-20&	03 25 56.3&	+76 16 38.4&	2.2 x 1.7&	$23.5 \pm 0.2$	&	face-on spiral\\
KKH01-21 &    	03 30 54.7  &    +81 33 49.1&     2.7 x 2.0&       $24.2 \pm 0.1$&  	Gal. neb\\
WHI J0337+75	&03 37 25.2&	+75 15 02.4&	2.0 x 1.0&	$23.2 \pm 0.2$	&	spiral\\
UGCA 86	&	03 59 48.3&	+67 08 18.6&	5.5 x 4.0&	$21.9 \pm 0.1$	&	irr. gxy nearby\\
WHI J0401+80 &   04 01 47.6     & +80 59 47.3&     1.4 x 0.5      & $24.0 \pm 0.1$&  	Gal. neb.\\
ZOAG G135.05+16.10&     04 03 31.5 &     +74 15 06.8&     1.3 x 1.4 &      $24.8 \pm 0.1$&	spiral?\\
ZOAG G135.23+16.04 &    04 04 44.6&      +74 05 39.8 &    0.8 x 1.1&       $24.1 \pm 0.1$ & 	Gal.neb?\\
WHI J0431+44&	04 31 30.5&	+44 24 13.0&	5.8 x 2.8&	$23.4 \pm 0.4$&		Gal. neb\\
WHI B0441+02&	04 43 44.4&	+02 59 34.9&	1.6 x 0.7&	$23.7 \pm 0.2$&		galaxy\\
ZOAG G167.44-04.85&	04 52 24.8&	+36 25 53.3&	1.4 x 1.0&	$23.4 \pm 0.4$&		spiral\\
KKH01-29    &	04 56 53.9 &     +37 57 09.6  &   1.7 x 1.2&       $23.3 \pm 0.1$  &	Neb + star cl\\
WHI J0512-00&	05 12 48.6&	-00 47 22.5&	2.4 x 1.7&	$24.5 \pm 0.3$	&	Gal. neb.\\
WHI J0514+55 &   05 14 33.4 &     +55 38 10.4&     1.6 x 0.8 &      $23.40 \pm 0.05$&	face-on spiral\\
WHI J0515+56  &  05 15 09.7&      +56 15 07.7 &    4.9 x 1.7&       $25.0 \pm 0.4$  &	Gal.neb.\\
WHI J0529+72&	05 29 17.7&	+72 27 13.8&	1.6 x 1.5&	$24.1 \pm 0.4$	&	galaxy\\
WHI J0620+49 &   06 20 18.6&      +49 19 18.2&     1.6 x 0.8&       $24.0 \pm 0.05$& 	galaxy\\
WHI J0623+09&	06 23 34.4&	+09 56 30.0&	1.5 x 1.1&	$23.2 \pm 0.2$	&	Gal. neb\\
KKH01-38    &	06 47 56.0    &  +47 30 40.6&     1.1 x 0.9     &  $24.5 \pm 0.2$&  	galaxy\\
ZOAG 211.04+01.19&	06 52 11.9&	+02 13 18.3&	1.5 x 1.1&	$22.8 \pm 0.2$&		Gal. neb\\
WHI J0706+58   & 07 06 45.8&      +58 46 25.8   &  1.8 x 1.0&       $25.0 \pm 0.3$  &	Gal.nebula\\
WHI J0711+14&	07 11 54.2&	+14 21 39.4&	5.0 x 1.5&	$24.1 \pm 0.4$	&	Gal. neb.\\
WHI J0734+20&	07 34 32.9&	+20 56 07.1&	1.4 x 1.0&	$24.4 \pm 0.4$	&	Gal. neb.\\
DDO 45	&	07 36 06.9&	+02 42 16&	2.5 x 1.6/4.3&	$23.4 \pm 0.1$	&	PN\\
WHI J0826+22&	08 26 01.8&	+22 10 06.4&	4.5 x 5.4&	$24.6 \pm 0.1$	&	Gal. neb.\\
WHI J0905+78&	09 05 47.6&	+78 00 50.4&	0.7 x 0.5&	$23.9 \pm 0.2$	&	ell. gxy\\
WHI J0910+73&	09 10 12.1&	+73 26 19.8&	1.9 x 1.8&	$24.2 \pm 0.4$	&	Sph. gxy nearby\\
WHI J0921+76&	09 21 10.5&	+76 25 30.2&	6.3 x 7.9&	$24.1 \pm 0.05$	&	Gal. neb.\\
KKH01-49&	09 21 57.6&	+50 16 14.7&	1.2 x 0.9&	$23.3 \pm 0.2$	&	dSph\\
WHI J0943+31&	09 43 38.7&	+31 59 25.4&	2.1 x 2.3&	$24.3 \pm 0.1$	&	Sph. gxy \\
KK98-77	&	09 50 15.0&	+67 30 42.9&	3.1 x 1.6&	$24.2 \pm 0.2$	&	diffuse el.\\
KK98-81  &   	09 57 02.9  &    +68 35 34.2&     4.4 x 1.5 &      $24.2 \pm 0.2$&		spheroidal\\
KKH01-57  &   	10 00 06.5 &     +63 10 58.2 &    0.7 x 0.7&       $25.3 \pm 0.2$ & 	res.spheroidal\\
DDO 71	&	10 05 08.8&	+66 33 30.0&	1.6 x 1.3&	$23.0 \pm 0.3$	&	Sph. gxy nearby\\
WHI J1007+67&	10 07 01.7&	+67 49 39.6&	2.2 x 0.8&	$23.7 \pm 0.2$	&	spheroidal \\
KKH01-59     & 	10 10 16.2 &     +62 54 49.7&     1.6 x 0.9&       $\ldots$           &  	res.spheroidal\\
FS90-005&	10 42 34.4&	+12 09 02.2&	1.2 x 0.8&	$24.2 \pm 0.2$	&	Sph. gxy\\
FS90-014&	10 46 25.0&	+14 01 30.6&	0.8 x 0.9&	$23.8 \pm 0.2$	&	Sph. gxy\\
FS90-021&	10 46 57.6&	+12 59 56.4&	0.8 x 0.9&	$24.2 \pm 0.2$	&	Sph. gxy\\
WHI J1048+20&	10 48 27.7&	+20 51 16.8&	1.8 x 1.7&	$24.9 \pm 0.2$	&	Gal. neb.(?)\\
KKH01-63&	10 48 38.2&	+82 25 36.5&	6.3 x 2.0&	$23.9 \pm 0.2$	&	Gal. neb.\\
UGCA 220&	10 49 23.2&	+64 43 13.0&	2.0 x 0.9&	$24.3 \pm 0.2$	&	Gal. neb.\\
DDO 87	&	10 49 38.1&	+65 31 46.9&	2.0 x 1.0&	$23.5 \pm 0.2$	&	dSph\\
WHI J1050+64a&	10 50 15.1&	+64 46 25&	1.9 x 2.3&	$24.8 \pm 0.2$	&	refl. neb.\\
KK98-96	&	10 50 27.1&	+12 21 32.6&	1.5 x 0.9&	$24.7 \pm 0.4$	&	Sph. gxy\\
WHI J1050+64b&	10 50 39.4&	+64 49 52&	2.6 x 1.3&	$24.5 \pm 0.5$	&	refl. neb.\\
KKH01-64&	10 51 32.2&	+03 27 21.6&	1.2 x 0.8&	$23.0 \pm 0.2$	&	Gal. neb./irr gxy?\\
WHI J1053+24&	10 53 07.3&	+24 55 22.3&	4.4 x 2.5&	$24.7 \pm 0.2$	&	Gal. neb.\\
KK98-108&	11 40 03.6&	+46 28 42.9&	0.6	&	$24.3 \pm 0.1$	&	Sph. gxy\\
PGC 36594&	11 44 54.2&	+02 09 48.6&	1.4 x 1.0&	$24.5 \pm 0.2$	&	Sph/irr\\
PGC 39058&	12 14 08.4&	+66 05 41&	1.9? x 0.7&	$\ldots$	&	res. gxy\\
DDO 113	&	12 14 57.9&	+36 13 03.5&	2.1 x 1.2&	$23.9 \pm 0.2$	&	Sph. gxy resolving\\
UGCA 275	&12 14 59.7&	+09 33 58.9&	1.7 x 1.1&	$23.1 \pm 0.2$	&	spiral (?) gxy\\
PGC 40640&	12 26 05.7&	+08 58 05.9&	1.4 x 1.0&	$23.2 \pm 0.2$	&	El. gxy\\
VCC 1052&	12 27 55.1&	+12 22 15.6&	2.1 x 1.9&	$25.1 \pm 0.2$	&	Sph. gxy\\
VCC 1287&	12 30 24.6&	+13 58 53.7&	2.0 x 1.5&	$24.5 \pm 0.2$	&	Sph. gxy\\
UGC 7673&	12 31 58.6&	+29 42 33.3&	1.0 x 0.9&	$23.5 \pm 0.2$	&	res. gxy\\
DDO 133	&	12 32 54.6&	+31 32 21.5&	2.8 x 1.6&	$23.6 \pm 0.3$	&	res. gxy\\
WHI J1233+15	&12 33 29.6&	+15 14 07.3&	1.0 x 0.8&	$25.3 \pm 0.4$	&	Sph. gxy\\
KDG 171	&	12 39 02.8&	-00 39 49.1&	1.2 x 0.5&	$24.1 \pm 0.2$	&	Sph. gxy resolving\\
PGC 42397&	12 39 53.3&	-00 28 41.1&	1.0 x 0.5&	$24.2 \pm 0.2$	&	Gal. neb (?)\\
KK98-166&	12 49 12.7&	+35 36 50.4&	0.9 x 0.8&	$25.4 \pm 0.5$	&	galaxy\\
PGC 43523&	12 51 11.1&	+11 14 38.2&	0.6	&	$24.2 \pm 0.2$	&	Sph. gxy\\
PGC 43654&	12 52 21.1&	+21 37 46.2&	2.7 x 1.2&	$23.3 \pm 0.1$	&	spiral(?) gxy\\
GR 66	&	12 56 25.0&	+15 05 09.1&	1.4 x 0.8&	$23.5 \pm 0.2$	&	Sph. gxy\\
LSBC F575-04&	13 04 30.0&	+17 45 33.9&	0.9 x 0.5&	$23.9 \pm 0.1$	&	elliptical gxy\\ 
WHI J1308+54&	13 08 30.9&	+54 37 58.9&	0.8 x 0.8&	$23.5 \pm 0.2$	&	Irreg. gxy\\
UGCA 337&	13 12 58.5&	+41 47 10.9&	1.3	&	$22.7 \pm 0.2$	&	Sph. gxy\\
WHI J1313+10	&13 13 47.4&	+10 03 10.4&	1.5 x 0.9&	$24.3 \pm 0.06$	&	Sph. gxy nearby\\
LSBC F650-01	&14 16 21.9&	+13 52 23.2&	1.6 x 1.7&	$24.8 \pm 0.2$	&	Gal. neb ?\\
WHI J1425+52&	14 25 32.3&	+52 35 17.7&	2.1 x 1.0&	$24.6 \pm 0.3$	&	Gal. neb.\\
WHI J1545+17&	15 45 43.8&	+17 18 50.3&	2.0 x 1.2&	$24.7 \pm 0.3$	&	galaxy\\
KKR99-25&	16 13 48.34&	+54 22 20.4&	1.2 x 0.7&	$23.9 \pm 0.2$	&	res. sph.\\
WHI J1627+11&	16 27 36.9&	+11 56 01.8&	2.7 x 1.3&	$24.6 \pm 0.3$	&	Gal. neb.\\
WHI J1633+86&	16 33 53.5&	+86 08 28.2&	0.8 x 0.6&	$23.2 \pm 0.3$	&	galaxy\\
WHI J1655+69&	16 55 36.6&	+69 55 38.8&	0.5 x 0.35&	$22.7 \pm 0.1$	&	spiral (?) gxy\\
WHI J1723+38&	17 23 48.0&	+38 50 30.1&	4.1 x 1.9&	$25.3 \pm 0.3$	&	Gal. neb.\\
WHI J1728+29&	17 28 10.5&	+29 27 32.5&	4.2 x 2.5&	$25.3 \pm 0.6$	&	Gal. neb.\\
LSB F520-2&	17 38 18.1&	+25 59 03.6&	1.2 x 0.4&	$24.4 \pm 0.2$	&	Spiral?\\
WHI J1745+17&	17 45 43.8&	+17 18 51.6&	1.7 x 1.2&	$24.6 \pm 0.3$	&	barred spiral gxy\\
WHI J1754+04&	17 54 21.6&	+04 10 47.4&	5.5 x 5.0&	$22.9 \pm 0.2$	&	Gal. neb\\
KKR99-40&	18 05 07.4&	+23 08 28.9&	1.4 x 0.8&	$25.4 \pm 0.2$	&	galaxy ?\\
WHI J1813+06&	18 13 18.2&	+06 46 40.1&	2.4 x 1.9&	$23.8 \pm 0.2$	&	Gal. neb.\\
WHI J1816+29&	18 16 24.9&	+29 49 26.3&	3.4 x 3.1&	$24.8 \pm 0.3$	&	Gal. neb. (PN?)\\
WHI J1824+24&	18 24 22.6&	+24 36 03.9&	2.6 x 2.8&	$24.9 \pm 0.3$	&	Gal. neb. (PN?)\\
WHI J1831+24&	18 31 19.5&	+24 55 39.0&	6.3 x 3.6&	$24.0 \pm 0.4$	&	Gal. neb.\\
CGMW 5-5772&	18 37 06.1&	+12 23 08.4&	1.4 x 1.0&	$22.7 \pm 0.2$	&	Spiral gxy\\
WHI J1844+28&	18 44 03.7&	+28 55 07.3&	3.0 x 1.4&	$25.1 \pm 0.2$	&	Gal. neb.\\
WHI J1856+52&	18 56 24.8&	+52 55 09.3&	2.1 x 1.8&	$24.1 \pm 0.3$	&	Gal. neb.\\
WHI J1859+45&	18 59 37.6&	+45 16 49.8&	3.5 x 2.1&	$24.0 \pm 0.3$	&	Gal. neb.\\
WHI J1909+50&	19 09 04.8&	+50 28 11.5&	5.9 x 2.4&	$24.1 \pm 0.2$	&	Gal. neb\\
WHI J1913+41&	19 13 34.3&	+41 09 30.3&	4.3 x 2.8&	$24.9 \pm 0.5$	&	Gal. neb.\\
WHI J1919+44 &    19 19 30.8&      +44 45 42.7&     3.1 x 2.4&	$25.3 \pm 0.3$	&	emission neb.\\
WHI J1932+08&	19 32 55.6&	+08 26 11.7&	0.7	&	$22.1 \pm 0.2$	&	face-on spiral\\
WHI J1933+55&	19 33 26.2&	+55 56 47.5&	2.6 x 2.0&	$24.3 \pm 0.2$	&	Gal. neb.\\
WHI J1945+22&	19 45 00.0&	+22 45 47.5&	3.5 x 4.0&	$23.7 \pm 0.3$	&	Gal. neb.\\
WHI J2004+64 &    20 04 54.6&	+64 36 18.8&	7.8 x 8.0&	$24.1 \pm 0.2$	&	Gal. neb.\\
WHI J2024+52&	20 24 29.2&	+52 49 05.0&	1.6 x 1.6&	$23.5 \pm 0.1$	&	Gal. neb.\\
WHI J2031+00&	20 31 15.5&	+00 08 22.1&	4.0 x 2.7&	$24.2 \pm 0.5$	&	Gal. neb.\\
ZOAG G093.12+08.90&	20 40 29.2&	+56 27 40.1&	1.7 x 1.6&	$23.6 \pm 0.2$&		Spiral gxy\\
KKR99-59&	21 03 25.3&	+57 16 44.9&	2.5 x 1.8&	$23.5 \pm 0.2$	&	Gal. neb.\\
WHI J2125+44&	21 25 52.3 &	+44 23 18.4&	0.8 x 0.7&	$22.0 \pm 0.2$	&	Barred spiral, or PN\\
WHI J2128+44&	21 28 26.8&	+44 39 56.7&	0.8	&	$21.9 \pm 0.4$	&	face-on spiral\\
WHI J2133+31&	21 33 42.6&	+31 17 20.0&	1.7 x 1.3&	$24.0 \pm 0.5$	&	Gal. neb.\\
WHI J2159+18&	21 59 36.6&	+18 14 50.5&	6.4 x 3.8&	$24.7 \pm 0.5$	&	Gal. neb.\\
WHI J2201+71&	22 01 38.3&	+71 46 35.1&	$>10$ x 8?&	$24.6 \pm 0.2$	&	Gal. neb.\\
WHI J2205+43&	22 05 22.6&	+43 49 26.5&	$>12$ x$>7$	&	$24.4 \pm 0.2$	&	Gal. neb.\\
KKR99-289.2&	22 11 45.2&	+45 36 42.5&	1.6 x 1.7&	$23.4 \pm 0.2$	&	barred spiral gxy\\
WHI J2219+20&    22 19 59.1 &     +20 20 23.4&     2.6 x 3.5 &      $25.0 \pm 0.2$	&	Gal. neb.\\
Sharpless 141&   22 28 37.9&      +61 37 55.3 &    4.2 x 4.8&       $23.3 \pm 0.05$	&	emission neb. \\
WHI J2234+20&	22 34 47.9&	+20 29 10.5&	4.9 x 3.0&	$25.6 \pm 0.4$	&	Gal. neb. \\
WHI J2259+58&	22 59 11.8&	+58 44 39.0&	5.7 x 6.0&	$22.7 \pm 0.3$	&	Gal. neb. \\
WHI J2309+44&	23 09 00.1&	+44 46 26.0&	2.6 x 1.6&	$24.05\pm 0.05$	&	Gal. neb. \\
WHI J2312+25&	23 12 11.0&	+25 07 51.3&	1.8 x 1.2&	$24.6 \pm 0.4$	&	Gal. neb. \\
WHI J2319+43 &   23 19 40.4 &     +43 47 19.3&     2.2 x 3.3 &      $24.2 \pm 0.1$ &		Gal.neb \\
WHI J2353+70  &  23 53 54.1&      +70 05 14.2 &    1.2 x 0.6&       $23.7 \pm 0.1$&		face-on spiral \\
\enddata
\tablecomments{Observational data on Local Group dwarf galaxy candidates in the northern
sky.  The list was produced
through visual examination of POSS-II survey plates; positions, sizes and surface brightness
measurements were obtained during follow-up observations with the Kitt Peak 2.1m telescope.
The identification as Galactic nebulosity, etc., is based on the morphology of the follow-up
image, supplemented by other information as available.  In two cases the surface brightness
could not be measured; these are shown by ``$\ldots$'' in the appropriate place in the table.}
\end{deluxetable}

\begin{deluxetable}{lcccl}
\rotate
\tablewidth{0pt}
\tablecaption{Southern Sky Local Group Dwarf Galaxy Candidates \label{southcand}}
\tablehead{
\colhead{Name} & \colhead{RA} & \colhead{Dec} & \colhead{Surface 
Brightness} & \colhead{Type} \\
& \colhead{(2000.0)} & \colhead{(2000.0)} &  \colhead{$R$ mag arc sec$^{-2}$} & }
\startdata
ESO 410G005	&00 15 31.5&	-32 10 52.5&	$22.5 \pm 0.2$&		nearby dSph\\
Cetus		&00 26 10.9&	-11 02 42.0&	$24.2 \pm 0.2$&		LG dwarf\\
Abell S143	&01 15 35.5&	-62 16 07.1&	$23.8 \pm 0.2$&		galaxy cluster\\
WHI B0200-03	&02 02 57.1&	-03 15 15.1&	$23.1 \pm 0.4$&		star cluster\\
ESO 298G033	&02 21 28.1&	-38 48 02.7&	$24.0 \pm 0.1$&		galaxy\\
PGC 9140	&02 24 43.7&	-73 30 50.3&	$23.7 \pm 0.2$&		galaxy\\
WHI B0240-07	&02 42 38.5&	-07 20 20.5&	$24.0 \pm 0.1$&		galaxy\\
KKH01-28	&04 43 44.5&	+02 59 45.8&	$23.8 \pm 0.1$&		irr gxy\\
ESO 85G088	&05 27 10.2&	-63 14 25.8&	$23.4 \pm 0.1$&		galaxy/Gal. neb.?\\
*WHI J0551-39	&05 51 33.2&	-39 59 02.9&	$25.2 \pm 0.5$&		Gal. neb, 4.2x1.2\\
WHI B0619-07	&06 22 13.7&	-07 50 25.8&	$22.7 \pm 0.2$&		galaxy (?)\\
WHI B0652+00	&06 54 36.4&	+00 14 55.3&	$24.0 \pm 0.1$&		Gal. neb\\
PGC 20125	&07 05 18.5&	-58 30 57.0&	$24.1 \pm 0.2$&		nearby galaxy\\
WHI B0713-44	&07 15 00.3&	-44 23 54.7&	$23.9 \pm 0.4$&		Gal. neb.\\
PGC 20635	&07 18 37.9&	-57 24 46.5&	$23.9 \pm 0.3$&		irr gxy\\
WHI B0717-07	&07 19 40.4&	-07 13 11.8&	$24.2 \pm 0.1$&		PN\\
ESO 368G004	&07 32 54.1&	-35 29 15.1&	$22.8 \pm 0.1$&		galaxy\\
PGC21406	&07 37 12.7&	-69 20 38.0&	$23.7 \pm 0.2$&		galaxy\\
WHI B0740-02	&07 43 21.6&	-02 32 12.6&	$23.7 \pm 0.1$&		gxy cluster\\
WHI B0744-05	&07 46 43.7&	-05 47 17.0&	$23.9 \pm 0.1$&		galaxy\\
KK00-24		&07 51 23.3&	-55 27 08.1&	$23.7 \pm 0.1$&		galaxy\\
MeWe 1-1	&08 53 36.2&	-54 04 54.7&	$24.7 \pm 0.2$&		PN\\
WHI B0921-36	&09 23 07.0&	-36 26 06.8&	$23.9 \pm 0.3$&		Gal. neb.\\
ESO 126G019	&09 34 14.0&	-61 16 57.5&	$22.2 \pm 0.1$&		galaxy\\
KDG 58		&09 40 26.7&	+00 02 45.1&	$25.2 \pm 0.6$&		Gal. neb/gxy? \\
WHI B0959-61	&10 00 28.2&	-62 08 52.3&	$23.2 \pm 0.3$&		Gal neb.\\
Antlia		&10 04 04.1&	-27 19 51.6&	$23.8 \pm 0.2$&		LG dwarf\\
*WHI J1019-23	&10 19 53.2&	-23 48 17.2&	$24.7 \pm 0.1$&		Gal. neb, 3.5x1.4\\
PGC 30367	&10 22 29.1&	-33 07 34.4&	$24.0 \pm 0.3$&		galaxy\\
WHI B1030-62	&10 32 19.9&	-63 09 57.5&	$23.4 \pm 0.3$&		Gal neb.\\
ESO 215G009	&10 57 30.3&	-48 10 32.1&	$22.7 \pm 0.2$&		galaxy\\
KKS00-23	&11 06 12.0&	-14 24 25.7&	$24.5 \pm 0.2$&		galaxy\\
KK00-41		&11 19 40.3&	-69 05 12.6&	$22.7 \pm 0.3$&		Gal. neb? gxy?\\
PGC 35171	&11 26 10.2&	-72 36 58.4&	$23.0 \pm 0.2$&		galaxy\\
*WHI J1129-13	&11 29 29.2&	-13 25 53.9&	$24.6 \pm 0.1$&		Gal. Neb, 3.5x4.0\\
PGC 36594	&11 44 54.1&	+02 09 52.7&	$24.5 \pm 0.1$&		galaxy\\
WHI B1241-54	&12 44 33.3&	-54 24 57.4&	$23.4 \pm 0.1$&		Gal. neb\\
WHI B1243-20	&12 45 41.2&	-20 31 41.8&	$24.3 \pm 0.1$&		Gal. neb\\
WHI B1249-33	&12 51 48.7&	-33 30 58.6&	$\ldots$&		not real? \\
ESO 269G066	&13 13 08.8&	-44 53 22.5&	$22.4 \pm 0.1$&		res. gxy\\
PGC 46680	&13 22 02.4&	-42 32 08.7&	$23.4 \pm 0.1$&		galaxy\\
PGC 48001	&13 36 11.7&	-56 32 21.3&	$22.2 \pm 0.1$&		Gal. neb/gxy (?)\\
PGC 48178	&13 38 10.3&	-56 28 43.2&	$21.7 \pm 0.2$&		barred spiral\\
ESO 174G001	&13 47 58.5&	-53 21 10.6&	$23.0 \pm 0.3$&		galaxy\\
MeWe 1-4	&14 17 32.2&	-52 26 24.6&	$24.1 \pm 0.5$&		PN\\
WHI B1425-47	&14 28 21.9&	-47 26 58.3&	$22.9 \pm 0.2$&		Galactic neb.\\
WHI B1432-16	&14 35 25.4&	-17 10 01.3&	$23.5 \pm 0.1$&		galaxy\\
WHI B1432-47	&14 35 50.6&	-47 58 17.5&	$23.7 \pm 0.1$&		Gal. neb\\
WHI B1444-16	&14 47 00.4&	-16 57 17.7&	$22.9 \pm 0.1$&		face-on spiral\\
KK00-61		&15 10 33.5&	-67 56 37.8&	$22.8 \pm 0.4$&		Gal neb.? gxy?\\
WHI B1517-41	&15 21 06.7&	-41 48 59.4&	$23.5 \pm 0.4$&		Gal. neb\\
KKS00-48	&16 05 40.4&	-04 34 20.1&	$23.7 \pm 0.1$&		galaxy\\
PGC 57387	&16 10 45.7&	-65 44 22.6&	$22.6 \pm 0.1$&		Gal neb.\\
WHI B1619-67	&16 24 59.1&	-67 07 31.8&	$23.0 \pm 0.1$&		Gal. neb\\
PGC 58179	&16 27 20.9&	-60 27 36.0&	$23.1 \pm 0.6$&		Gal. neb? galaxy? \\
WHI B1728-08	&17 31 29.2&	-08 19 14.5&	$24.1 \pm 0.2$&		PN\\
WHI B1751-07	&17 53 49.3&	-07 03 03.7&	$23.3 \pm 0.1$&		spiral galaxy\\
*WHI J1828-52	&18 28 30.1&	-52 48 44.0&	$24.3 \pm 0.2$&		Gal. neb, 4.5x4.6\\
PGC 62147	&18 37 24.4&	-57 25 50.5&	$23.9 \pm 0.2$&		galaxy\\
ESO 458G011	&18 59 32.3&	-31 12 43.5&	$22.1 \pm 0.3$&		spiral?\\
WHI B1919-04	&19 22 01.8&	-04 12 05.5&	$23.4 \pm 0.1$&		spiral galaxy\\
WHI B1952-04	&19 55 39.5 &	-04 23 45.4&	$23.6 \pm 0.1$&		barred spiral\\
ESO 027G002	&21 52 18.7&	-80 34 23.9&	$23.6 \pm 0.2$&		galaxy\\
WHI B2212-10	&22 15 26.3&	-10 28 33.4&	$24.2 \pm 0.1$&		galaxy\\
ESO 468G020	&22 40 44.1&	-30 48 02.1&	$23.5 \pm 0.2$&		nearby dSph\\
SC 2		&23 20 35.3&	-31 54 34.0&	$24.3 \pm 0.2$&		face-on spiral?\\
UKS2		&23 26 27.4&	-32 23 12.4&	$22.2 \pm 0.1$&		res. galaxy\\
\enddata
\tablecomments{Observational data on Local Group dwarf galaxy candidates in the
southern sky.  The list was produced through visual examination of SRC/ESO survey plates;
positions and surface brightness measurements were obtained during follow-up
observations with the CTIO 1.5m and INT 2.5m telescopes.  The identification as
Galactic nebulosity, etc., is based on the morphology of the follow-up
image, supplemented by other information as available.  In one case the surface brightness
could not be measured; this is shown by ``$\ldots$'' in the appropriate place in the table.}
\end{deluxetable}

\begin{deluxetable}{lccllr}
\rotate
\tablewidth{0pt}
\tablecaption{Northern Sky Local Group Dwarf Non-candidates \label{northnon}}
\tablehead{
\colhead{Name} & \colhead{RA (2000)} & \colhead{Dec (2000)} & \colhead{Type} & \colhead{Alternate Names}
& \colhead{Heliocentric RV, km s$^{-1}$}}
\startdata
UGC 12894       &00 00 22.6 &     +39 29 44 &      galaxy       &   LEDA 35		&	335  \\
A66-86		&00 01 31.0&	+70 42 30&	PN		&PK 118+08 2&\\
Andromeda III	&00 35 22.7&	+36 30 17&	LG galaxy	&KK98-5	&\\
Andromeda I	&00 45 39.8&	+38 02 28&	LG galaxy 	&KK98-8&\\
PHL 932		&00 59 56.7&	+15 44 14&	PN		&PK 125-47 1&\\
Andromeda II	&01 16 30&	+33 25.9&	LG galaxy	&KK98-12&\\
UGC 672		&01 06 17.9&	+44 57 15&	galaxy		&TC 12	&		708  \\
DDO 9		&01 10 44.0&	+49 36 08&	galaxy		&UGC 731&			639  \\
IS96 0110+0046	&01 12 50.7&	+01 02 49&	galaxy		&$\ldots$&			1105  \\
UGC 1084	&01 31 22.1&	+23 57 14&	galaxy		&LEDA 5664&		3414  \\
DDO 13		&01 40 10.4&	+15 54 19&	galaxy		&UGC 1176&		633  \\
KDG 10		&01 43 37.2&	+15 41 43&	galaxy		&LEDA 6354&		791  \\
ZOAG G131.13-06.38	&01 49 29.6&	+55 34 07&	galaxy	&LEDA 166411&		640  \\
Cas 1		&02 06 02.8&	+68 59 59&	LG galaxy	&KK98-19, ZOAG G129.56+07.09&	35  \\
ZOAG G135.74-04.53&	02 24 34.3&	+56 00 39&	galaxy	&KKH01-11&		310  \\
A66-6		&02 58 41.9&	+64 30 06&	PN		&PK 136+04 1&\\
PN G136.3+05.5	&03 03 48.8&	+64 53 28&	PN		&$\ldots$&\\
HaWe 2		&03 11 01.3&	+62 47 45&	PN		&PK 138+04 1, HDW 2&\\
UGC 2767	&03 35 33.2&	+80 05 08&	galaxy		&ZOAG G129.82+19.49&	2210  \\
UGCA 86		&03 59 50.5&	+67 08 37&	galaxy		&PGC 14241	&	67  \\
ZOAG G145.42+07.36&	04 16 39.7&	+60 55 33&	galaxy	&LEDA 89976	&	1109  \\
EGB 3		&04 25 16.3&	+72 48 21&	PN		&PK 137+16 1, Cam A&\\
EGB 0427+63	&04 32 04.9&	+63 36 49&	LG galaxy	&UGCA 92, PGC 15439&	-99  \\
Cam B		&04 53 07.7&	+67 06 01&	galaxy		&HKK L41	&		75  \\
HS98 OD		&05 29 17.5&	+72 27 11&	galaxy		&$\ldots$	&		1089  \\
DDO 38		&05 33 37.5&	+73 43 26&	galaxy		&UGC 3317	&	1240  \\
DDO 39		&05 56 36.0&	+75 19 02&	galaxy		&UGC 3371	&	816  \\
A66-16		&06 43 55.5&	+61 47 25&	PN		&PK 153+22 1&\\
UGC 3817	&07 22 44.5&	+45 06 31&	galaxy		&MCG+08-14-000	&	438  \\
DDO 44		&07 34 11.5 &	+66 52 47&	galaxy		&UGCA 133, KK98-61&	\\
DDO 46		&07 41 26.0&	+40 06 40&	galaxy		&UGC 3966	&	361  \\
HaWe 10		&07 55 11.3&	+09 33 09&	PN		&HDW 7&\\
M81 dwA		&08 23 55.1&	+71 01 56&	galaxy		&LEDA 139073	&	113  \\
DDO 52		&08 28 28.4&	+41 51 24&	galaxy		&UGC 4426	&	397  \\
DDO 53		&08 34 07.3&	+66 10 55&	galaxy		&UGC 4459, Zw VII 238&	19  \\
A66-28		&08 41 35.6&	+58 13 48&	PN		&PK 158+37 1&\\
KK98-69		&08 52 50.8&	+33 47 52&	galaxy		&$\ldots$&			489  \\
LSK86-84	&08 54.0&		+78 16&		galaxy	&$\ldots$&			1476  \\
A66-31		&08 54 13.2&	+08 53 53&	PN		&PK 219+31&\\
UGC 4683	&08 57 54.4&	+59 04 58&	galaxy		&MCG+10-13-046	&	928  \\
VLSB F564-V03	&09 02 53.7&	+20 04 30&	galaxy		&ESDO 564-08	&	481  \\
UGC 4945	&09 22 26.6&	+75 45 59&	galaxy		&UGCA 158	&	659  \\
HS98-103	&09 50 10.5&	+67 30 24&	galaxy		&KK98-77&\\
EGB 6		&09 52 59.0&	+13 44 35&	PN		&PK 221+46 1&\\
KDG 61		&09 57 03.1&	+68 35 31&	galaxy		&KK98-81, PGC 28731&	-135  \\
DDO 71		&10 05 06.2&	+66 33 31&	galaxy		&UGC 5428	&	\\
KDG 64		&10 07 01.7&	+67 49 38&	galaxy		&UGC 5442&\\
Leo I		&10 08 28.1&	+12 18 23&	LG galaxy	&UGC 5470, DDO 74	&	\\
UGC 5455	&10 08 50.2&	+70 38 03&	galaxy		&Mailyan 51	&	1291  \\
DDO 78		&10 26 28.0&	+67 39 35&	galaxy		&KK98-89	&		2550  \\
BK 6N		&10 34 29&	+66 00.5&	galaxy 		&KK98-91&\\
DDO 86		&10 44 30.1&	+60 22 05&	galaxy		&UGC 5846&		1022  \\
KDG 73		&10 52 57.1&	+69 32 57&	galaxy		&PGC 32667&		116  \\
KKH01-64	&10 51 32.0&	+03 27 14&	galaxy		&$\ldots$&			1070  \\
ISI96 1050+0245	&10 53 03.1&	+02 29 37&	galaxy		&KKS00-58&		1054  \\
UGC 6113	&11 02 48.6&	+52 06 59&	galaxy		&LEDA 33346&		951  \\
Leo B		&11 13 28.1&	+22 09 10&	LG galaxy	&UGC 6253&\\
DDO 97		&11 48 57.2&	+23 50 16&	galaxy		&UGC 6782&		525  \\
UGCA 259	&11 58 52.8&	+45 43 55&	galaxy		&KK98-116&		1154  \\
DDO 113		&12 14 57.9&	+36 13 07&	galaxy		&UGCA 276&		284  \\
VCC 169		&12 15 56.4&	+09 38 56&	galaxy		&GRDG +09 5&		2222  \\
UGC 7307	&12 17 04.5&	+10 00 19&	galaxy		&LEDA 39380&		1184  \\
DDO 131		&12 31 58.6&	+29 42 34&	galaxy		&UGC 7673&		642  \\
DDO 133		&12 32 54.6&	+31 32 31&	galaxy		&UGC 7698&		331  \\
UGCA 285	&12 33 08.0&	-00 32 01&	galaxy		&IDI96 1230-0015	&	3279  \\
UGCA 292	&12 38 40.1&	+32 46 01&	galaxy		&LEDA 42275	&	307  \\
DDO 143		&12 44 25.1&	+34 23 12&	galaxy		&UGC 7916, I Zw 42&	607  \\
DDO 147		&12 46 59.7&	+36 28 34&	galaxy		&UGC 7949	&	333  \\
UGC 7995	&12 50 00.2&	+78 23 05&	galaxy		&Mailyan 80	&	1799  \\
NGC 4789A	&12 54 05.5&	+27 08 55&	galaxy		&DDO 154, UGC 8024	&376  \\
LoTr 5		&12 55 33.7&	+25 53 31&	PN		&PK 339+88 1&\\
KDG 215		&12 55 41.4&	+19 12 34&	galaxy		&LEDA 44055&		\\
DDO 165		&13 06 24.9&	+67 42 25&	galaxy		&UGC 8201, VII Zw 499	&37  \\
DDO 175		&13 25 29.3&	+57 49 18&	galaxy		&UGC 8441	&	1519  \\
DDO 181		&13 39 53.8&	+40 44 25&	galaxy		&UGC 8651	&	201  \\
DDO 187		&14 15 56.7&	+23 03 16&	galaxy		&UGC 9128	&	154  \\
UGC 9381	&14 34 34.7&	+36 17 17&	galaxy		&LEDA 52088	&	3028  \\
Palomar 5	&15 16 05.3&	-00 06 41&	globular cl.	&UGC 9792	&\\
UGC 9938	&15 37 12.0&	+30 04 37&	galaxy		&LEDA 55621	&	1865  \\
KKR99-22	&15 45 44&	+17 18.6&	galaxy		&LSB F583-5	&	3261  \\
UGC 10031	&15 45 45.7&	+61 33 21&	galaxy		&Mailyan 95 	&	898  \\
DDO 202		&15 51 15.1&	+16 19 46&	galaxy		&UGC 10061	&	2080  \\
Palomar 15	&17 00 02.4&	-00 32 31&	globular cl.	&C 1657-004&\\
UGC 10792	&17 14 01.6&	+75 12 13&	galaxy		&PGC 59888&		1233  \\
Draco		&17 20 12.4&	+57 54 55&	LG galaxy	&DDO 208&\\
A66-61		&19 19 10.2&	+46 14 52&	PN		&PK 077+14 1&\\
WeSb 5		&20 01 42.0&	+19 54 41&	PN		&PK 058-05 1&\\
UGC 11926	&22 09 31.1&	+18 40 54&	galaxy		&MGC+03-56-015&		1653  \\
KKR99-70	&22 28 05&	+23 22.0&	galaxy		&LSB F533-1&		1278  \\
DDO 213		&22 34 10.9&	+32 51 41&	galaxy		&UGC 12082&		802  \\
Palomar 13	&23 06 44.5&	+12 46 19&	globular cl.	&UGCA 435& \\
Andromeda VI	&23 51.7	&	+24 36	&	LG galaxy	&Pegasus dSph&\\
\enddata
\tablecomments{Objects selected as possible Local Group dwarf galaxies based on their morphology
on POSS-II survey plates, but rejected for various reasons.  
In the last column is the NED-derived radial velocity, which
explains rejection for most of the galaxies here listed.}
\end{deluxetable}

\begin{deluxetable}{lccllr}
\rotate
\tablewidth{0pt}
\tablecaption{Southern Sky Local Group Dwarf Non-candidates \label{southnon}}
\tablehead{
\colhead{Name} & \colhead{RA (2000)} & \colhead{Dec (2000)} & \colhead{Type} & \colhead{Alternate Names}
& \colhead{Heliocentric RV, km s$^{-1}$}}
\startdata
Sculptor	&01 00 09.4&	-33 42 32&	LG galaxy	&ESO 351G030&\\
ISI96 0110+0046	&01 12 50.7&	+01 02 49&	galaxy		&$\dots$&				1105  \\
Fornax		&02 39 59.3&	-34 26 57&	LG galaxy	&ESO 356G004&\\
DDO 27		&02 40 23.4&	+01 13 43&	galaxy		&UGC 2162&			1185  \\
UGCA 44		&02 49 22.2&	-02 39 14&	galaxy		&KKS00-52&			1094  \\
Lo 1		&02 56 58.4&	-44 10 18&	PN		&ESO 247G013, PK 255-59 1&\\
UGCA 65		&03 18 43.3&	-23 46 55&	galaxy		&AM 0316-235, ESO 481G019&	1535  \\
KDG 38		&03 23.9 &	-19 16	&	galaxy		&SGC 0321.1-1929	&		1545  \\
Horologium	&03 59 15.2&	-45 52 14&	galaxy		&ESO 249G036, AM 0357-460&	901  \\
PK 215-30 1	&05 03 07.5&	-15 36 23&	PN		&A55 6, PN A66 7&\\
DDO 234		&06 15 19.4&	-26 34 32&	galaxy		&AM 0613-263, UGCA 122	&	1800  \\
Carina		&06 41 36.7&	-50 57 58&	LG galaxy	&ESO 206G20A&\\
ESO 430G001	&07 55 12.4&	-28 09 58&	galaxy		&SGC 075310-2801.0		&1691  \\
ESO 561G002	&07 55 25.9&	-21 20 29&	galaxy		&SGC 075315-2112.5	&	922  \\
PK 224+15	&08 06 46.5&	-02 52 35&	PN		&K1-13, A66-25&\\
PGC 022808	&08 07 30.2&	-27 30 32&	galaxy		&SGC 080526-2721.8&		920  \\
ESO 371G030	&09 00 18.8&	-34 04 46&	galaxy		&AM 0858-335, SGC 085817-3353.0&	1338  \\
Pyxis		&09 07 57.8&	-37 13 17&	globular cl.	&C J0908-373&\\
DDO 57		&09 11 20.1&	-15 02 54&	galaxy		&MCG-02-24-001&			2049  \\
Palomar 3	&10 05 31.0&	+00 04 15&	globular cl.	&Sextans C, UGC 5439&\\
Lo 5		&11 13 54.2&	-47 57 01&	PN		&ESO 215-35, PK 286+11 1&\\
UGCA 285	&12 33 08.0&	-00 32 01&	galaxy		&ISI96 1230-0015	&		3279  \\
NGC 4942	&13 04 18.9&	-07 38 54&	galaxy		&IC 4136		&		1751  \\
DDO 163		&13 05 14.4&	-07 53 24&	galaxy		&MCG-01-33-082		&	1123  \\
DDO 195		&14 38 54&	-08 37.8&	galaxy		&LEDA 52345		&	1824  \\
Palomar 5	&15 16 05.3&	-00 06 41&	globular cl.	&UGC 9792&\\
Lo 9		&15 42 13.3&	-47 40 46&	PN		&PK 330+05 1&\\
PK 329-01 1	&15 54 50.9&	-51 22 35&	PN		&AM 1551-511, ESO 225G003&\\
Terzan 3	&16 28 40.1&	-35 21 13&	globular cl.	&ESO 390G006&\\
MeWe 1-11	&17 52 47.1&	-46 42 02&	PN		&PN G345.3-10.2&\\
PK 332-16.1	&17 54 09.6&	-60 49 58&	PN		&HaTr 7&\\
Lo 17		&18 27 50.0&	-37 15 52&	PN		&ESO 395G007, PK 356-11 1&\\
ESO 184G018	&19 09 47.8&	-55 35 11&	PN		&Lo 18, PK 341-24.1&\\
Arp 2		&19 28 44.1&	-30 21 14&	globular cl.	&C 1925-304&\\
Sag DIG		&19 29 59.0&	-17 40 41&	LG galaxy	&ESO 594-4, UKS 1927-17.7&\\
Aquarius	&20 46 51.8&	-12 50 52&	LG galaxy	&DDO 210&\\
ESO 288G40	&22 06 33.3 &	-42 51 27&	galaxy		&SGC 220330-4306.2	&	2212  \\
ESO 238G005	&22 22 30.4&	-48 24 13&	galaxy		&AM 2219-483, KK98-257	&	706  \\
ESO 238G016	&23 33 46.6&	-48 01 24&	galaxy		&AM 2230-481, SGC 223045-4816.9&	8300  \\
Tucana		&22 41 49.6&	-64 25 10&	LG galaxy	&SGC 223828-6440.9&\\
\enddata
\tablecomments{Objects selected as possible Local Group dwarf galaxies based on their morphology
on ESO/SRC survey plates, but rejected for various reasons.
In the last column is the NED-derived radial velocity, which
explains rejection for most of the galaxies here listed.}
\end{deluxetable}

\clearpage

\subsection{Notes on Particular Objects}

The following notes are presented in RA order for the northern objects, then
a few pertaining to the south are appended.

{\bf Northern Candidates:}

WHI J0004+52: M-shaped Galactic nebulosity.

WHI J0018+65: Planetary nebula, confirmed by H$\alpha$ image.

WHI J0039+66: Galactic nebula, with H$\alpha$ emission that looks like a bow shock.

WHI J0040+66: Galactic nebulosity.  Very ill-defined, to the point that the size
is a guess, and the surface brightness also. 

WHI J0144+17:  A faint bit of galactic nebulosity.  No apparent H$\alpha$.

WHI J0156+63: Arrowhead-shaped bit of Galactic nebulosity.

WHI J0156+13: A reflection nebula in the glare of HD 11861, to which it may be
related.  Blue in color with no apparent H$\alpha$.  We present an image below.

ZOAG G133.63-03.62: Face-on spiral galaxy.

WHI J0226+52: A very large Galactic nebulosity, extending off the field to the
north and east. 

KK98-20: Galaxy, elliptical overall but showing a trace a spiral structure.
\citet{HKKE00} reported a detection in HI at 760 mJy, with a systemic velocity
of -70 km/s, and concluded that it is a Galactic HI cloud.  However,
\citet{RS00} found a signal at 3.7 mJy showing 3994 km/s.  The latter detection
makes more sense together with the compact, spiral morphology that we see.
\citet{HKKE00} searched only out to 3970 km/s, which would explain how they missed
the extragalactic signal.  What they found at -70 km/s, however, is
not explained; there is no obvious Galactic object nearby, and \citet{RS00}
did not report it.

ZOAG G134.31+06.24: Bright face-on spiral.

ZOAG 139.32+04.85: Cam C, identified as an emission nebula by \citet{KSD03}.  
The ZOAG catalog seems to have listed each of the
two bright spots separately.  Pretty in H-alpha; probably a bipolar PN.

KKH01-20: Heavily extincted face-on spiral.  Identified as a possible dE by
\citet{HFL95}, from plate material which did not go deep enough to show the
spiral arms.

KKH01-21: Galactic nebulosity, with a sort of milky appearance.

WHI J0337+75: Spiralish galaxy, slightly lumpy.

UGCA 86: A nearby galaxy \citep{KSD03}.  Not faint; included because of
resolution on the survey plate.  Significant areas of H$\alpha$ emission.

WHI J0401+80: Galactic nebulosity; the figures in the table refer just to 
the brightest knot.  There is more spread
throughout the whole frame.

ZOAG G135.05+16.10: Probably a heavily extincted, face-on spiral.

ZOAG G135.23+16.04: A lumpy bit of nebulosity?  (Though it could  possibly be a heavily
extincted galaxy).

WHI J0431+44: Galactic nebulosity.

WHI B0441+02: A lumpy galaxy; it appeared in \citet{WHI02}.

ZOAG G167.44-04.85: Heavily extincted spiral.

KKH01-29: Galactic nebulosity, involved with an apparent star cluster.  The
stars make it even more difficult than normal to get a good surface
brightness.

WHI J0512-00: Galactic nebulosity (superimposed on some distant galaxies).

WHI J0514+55:  Nearly face-on spiral.

WHI J0515+56: Curve of galactic nebulosity.

WHI J0529+72:  Of irregular shape, from the morphology possibly an irregular 
galaxy or a Galactic reflection
nebula.  The latter initially appeared more likely in light of the lack of H-alpha
emission.
\citet{HS98} find a redshift of 1089, so it's probably the former.

WHI J0620+49: Wispy nebulosity, against a background of distant galaxies.  The
center is mottled, though, in a way that suggests a dwarf galaxy possibly nearing
resolution.

WHI J0623+09: Detected by IRAS.  Catalogued and analyzed by \citet{SSW96} as 
a galaxy in the Zone of Avoidance, and listed as such in on-line 
databases.  However, it was not found in HI by \citet{PAG97}; and was
detected in a CS line at 98 GHz by \citet{BNM96} at a redshift of
35 km/s.  Those data together with its morphology clearly  
identify it as Galactic nebulosity, apparently associated
with a group of stars.

KKH01-38: Probably a face-on barred spiral.  Bluish in color, with no apparent
H-alpha emission.

ZOAG 211.04+01.19: Catalogued as a galaxy, but it looks more like a reflection nebula.
It appears as a Zone of Avoidance galaxy in \citet{SSW96} and is
analyzed as such, but is clearly part of a molecular cloud complex in
\citet{BW94}.

WHI J0706+58: Nebula involved with three brightish stars.  They no doubt contribute to
the measured surface brightness, making it an upper limit.

WHI J0711+14: Another wisp of faint nebulosity.

WHI J0734+20: A wisp of very faint nebulosity.

DDO 45: Nebulosity, within a much fainter outer shell (25.4 mag sec$^{-2}$).
Catalogued as a galaxy,
and intended to be analyzed as such by \citet{SPE97}, but not confirmed
by HI and thus not included.  Shown by \citet{KP85} to be
a planetary nebula.

WHI J0826+22:  The central, brighter section of a very faint, wispy nebulosity.  The
size given is more than usually uncertain.  The position is for the center
of the brighter section; the fainter envelope extends well to the north and
east.

WHI J0905+78:  A slightly irregular elliptical galaxy, in the foreground of the
cluster Abell 719.

WHI J0910+73: Diffuse, faint spheroidal galaxy.  It appears to be on the verge of
resolving.

WHI J0921+76:  A very large, very faint swirl of nebulosity.  The position given is
for the center of the brighter swirl; it extends far to the south and a
bit more to the east than the west.

KKH01-49: Spheroidal, a bit grainy but not resolving.

WHI J0943+31:  A diffuse bit of light near NGC 2970 (which is {\em much} brighter).  It
appears to extend across the brighter galaxy; but its color is very red,
not much like 2970 at all, while the extension to the southwest is different again.
Probably a superposed dwarf, not related to the brighter galaxy.

KK98-77: Very diffuse elliptical galaxy.

KK98-81: Spheroidal nucleus with  very extended elliptical halo.

KKH01-57: Spheroidal, starting to resolve.

DDO 71: Nearby galaxy (M 81 group).

WHI J1007+67: Spheroidal galaxy.

KKH01-59: Spheroidal with bright star superimposed.  Photometry is impossible.

FS90-005: Fairly concentrated spheroidal.

FS90-014: Fairly concentrated spheroidal.

FS90-021: Small spheroidal.

WHI J1048+20: Probably a lump of Galactic nebulosity (though a very LSB spheroidal
cannot be ruled out).

KKH01-63: A wisp of Galactic nebulosity.  The position given is the center of
the brightest knot; it fades away far to the east and southeast.

UGCA 220: A wisp of Galactic nebulosity.

DDO 87: Spheroidal, starting to resolve.

WHI J1050+64: Galactic nebulosity, detected on the plates and here catalogued in two
parts, but undoubtedly connected.  The situation in this region warrants a short
discussion.

\citet{S76} noted a network of Galactic nebulosity near this area, though not at this
position, in spite of the high latitude (above $45\arcdeg$).  \citet{BKK84} catalogued 
WHI J1050+64b as BKK 7, probably a galaxy, since there was
no nebulosity visible nearby on the POSS-I prints.  Based on size and morphology
they considered it most probable that BKK 7 and similar objects in the area constituted
a newly-fragmenting protogalaxy at the distance of the M81 group, and derived figures
for its mass and luminosity.  \citet{BBP98} thought these objects most likely Galactic
cirrus in light of IRAS data, but could reach no definite conclusion.
From the images presented here, WHI J1050+64 is clearly a network of Galactic 
nebulosity, aligned with UGC 5932 by chance.  It has the morphology of
a turbulent fluid, no discernable H$\alpha$ emission and a blue color.

We draw two immediate conclusions from this matter: first, that classification by morphology
alone of objects near the threshold of detection is likely to be inaccurate; and second, at
the low surface brightness levels dealt with here, Galactic emission may be met in any
part of the sky.  We also recommend this region as a testing ground for any automatic
algorithms for the detection of faint objects.  We suggest that an algorithm which does not
detect parts of WHI J1050+64 is not sensitive enough, while an algorithm which could
identify them as Galactic cirrus would be very useful.

KK98-96: Diffuse; spheroidal?

KKH01-64: Galactic nebula, or irregular galaxy?  From morphology alone we can reach no
conclusions.

WHI J1053+24: Very faint Galactic nebulosity.

KK98-108: The regularity of its shape points to being a spheroidal galaxy; there are
wispy appendages to the south and west in our follow-up image which
are probably reflections from stars
in the field (they have a vaguely doughnut shape).  Our position differs
from that of NED by $0.1'$, which is a reasonable estimate of how much the
perceived center of a faint, diffuse object might vary between observers.

PGC 36594: Faint galaxy, spheroidal or possibly irregular.  Subsequent to our observations
a radial velocity of 1013 km s$^{-1}$ has appeared in NED.

PGC 39058:  A resolving galaxy almost under a bright (8th mag.) star.  H$\alpha$ shows that
the brightest knots are emission nebulae; but other resolved objects are
probably stars.  Photometry is impossible.

DDO 113: Spheroidal galaxy, starting to resolve.

UGCA 275: Galaxy; brightish nucleus, with very faint spiral arms.

PGC 40640: Spheroidal galaxy, a bit lumpy.

VCC 1052: Very diffuse, faint, round galaxy.

VCC 1287: Spheroidal galaxy, near to resolution.

UGC 7673:  Resolving.  A radial velocity of 644 km s$^{-1}$ has appeared in NED subsequent
to our observations.

DDO 133: Also resolving, with a radial velocity of 328 or 331 km/s.  Plenty of H$\alpha$
emission.

WHI J1233+15:  Faint spheroidal galaxy. 

KDG 171: Spheroidal galaxy, starting to resolve.

PGC 42397: A wisp of Galactic nebulosity, or possibly the bar of a faint galaxy,
or an irregular galaxy.

KK98-166: Galaxy; maybe very LSB barred spiral?

PGC 43523: Small spheroidal galaxy.

PGC 43654: Spiralish galaxy with extended faint envelope.  H$\alpha$ emission, especially,
seems to want to go in a spiral pattern.  The brightest knots are HII
regions; but there also seem to be a few supergiants.

GR 66: Spheroidal galaxy, not showing any desire to resolve.

LSBC F575-04:  Elliptical galaxy, elongated, with a superimposed star or nucleus.

WHI J1308+54: Irregular galaxy (or maybe barred spiral with an appendage). 
No detectable H$\alpha$.

UGCA 337: Spheroidal, just too far away to resolve.

WHI J1313+10: Elliptical/spheroidal galaxy (with three inconvenient stars
superimposed), starting to resolve; but the lack of clear resolution
with 2700s of $I$ on the 2.1m means it's well beyond the Local Group.

LSBC F650-01: Possibly Galactic nebulosity?  Also possibly a LSB spheroidal.

WHI J1425+52:  Galactic nebulosity; the position given is for the center of the bright
wisp, though a fainter extension goes well to the east.

WHI J1545+17: Spiralish; LSB arms around a bright nucleus?

KKR99-25: Resolving dwarf spheroidal, with an inconveniently placed bright star.

WHI J1627+11: Faint Galactic nebula.

WHI J1633+86: Lumpy galaxy.

WHI J1655+69: The faint halo of a small spiral (?) galaxy.  Although overall the
surface brightness is high (for our objects), the arms are almost a magnitude fainter.

WHI J1723+38: A faint wisp of Galactic nebulosity.

WHI J1728+29: Very faint nebulosity.

LSB F520-2: A barred spiral, seen almost edge-on?

WHI J1745+17: Barred spiral with very low surface brightness arms.  Brightness
measurements do not include the superposed star or nucleus.

WHI J1754+04: Galactic nebulosity.

KKR99-40: Possibly a faint galaxy (though it lies
at low Galactic latitude, there are
other galaxies visible nearby), but could also be Galactic nebulosity.

WHI J1813+06: Galactic nebulosity.

WHI J1816+29:  A faint ring (old PN?) of Galactic nebulosity.

WHI J1824+24: Galactic nebulosity.  Measurements are given for a ringlike structure,
connected to more wisps going off the edge of the field.  This is conceivably
a PN.

WHI J1831+24: Mottled Galactic nebulosity with distant galaxies in the background.

CGMW 5-5772: Multiarmed spiral galaxy, behind a lot of Galactic stars.

WHI J1844+28: A wisp of Galactic nebulosity.  There is more in the field as well as
leading out of it; this is the most coherent, compact part.

WHI J1856+52: Very faint, two wisps of nebulosity forming a part of a circle.  Although
it is not much (if any) brighter than flat-fielding residuals, observations at two
different observing runs give the same shape and surface brightness.

WHI J1859+45: A few wisps which might outline a larger area of galactic nebulosity.

WHI J1909+50: Galactic nebulosity.

WHI J1913+41: The brightest bit of Galactic nebulosity which just about fills the field.
The main uncertainty in surface brightness comes from not knowing what is sky
and what is fainter nebulosity.

WHI J1919+44: Very nice bipolar PN. 

WHI J1932+08: Face-on spiral, with a central bar (accentuated by a guiding error
in our follow-up image).

WHI J1933+55: A roundish piece of nebulosity.

WHI J1945+22: Large, faint nebulosity.  Due to the high star density, the surface brightness measurements
are even more uncertain than usual.

WHI J2004+64: Swirls of Galactic nebulosity. Little or no H$\alpha$.

WHI J2024+52: A roundish bit of Galactic nebulosity.  Crowded field. 

WHI J2031+00: Oval bit of Galactic nebulosity.

ZOAG G093.12+08.90: A face-on, extincted spiral galaxy.  There are wisps of
Galactic nebulosity within a few arc minutes.

KKR99-59: A diffuse, oval object, catalogued by \citet{KKR99} as a probable nearby
dwarf galaxy.  Its morphology here, together with the fact that it has
apparently not been seen in HI by \citet{HKK00} nor in H$\alpha$ by \citet{MKB03},
lead us to believe it to be Galactic reflection nebulosity.

WHI J2125+44: Bright (and near a bright star).  Probably a barred spiral, but
possibly a PN.

WHI J2128+44: A face-on spiral galaxy.

WHI J2133+31: Galactic nebulosity, which appears (though under
conditions which were anything but photometric) to be fainter in V than in R.

WHI J2159+18: One end of a large, faint nebulosity.

WHI J2201+71: Large and ill-defined nebulosity; no meaningful dimensions can be given.

WHI J2205+43: An enormous skein of Galactic nebulosity, extending off the frame to the
east, west, and north.

KKR99-289.2:  A barred multiarmed spiral.  Lumpy.  Catalogued by \citet{KKR99}; given a radial
velocity of 1145 km/s by \citet{HKK00}.  From the H$\alpha$ image most of the
lumps are HII regions, though not all.

WHI J2219+20: Galactic nebulosity.  The size and position given correspond to a
slightly brighter section, but there is more throughout almost the
whole field.  No H$\alpha$ emission detected.

Sharpless 141: Brightish emission nebula.  It is very difficult to get the surface
brightness of the nebula alone, since the star density is so high.  Catalogued by 
\citet{S59} as an HII region, though this reference was apparently too old to have
been entered into Simbad when we checked.

WHI J2234+20:  Large, faint Galactic nebulosity; no apparent H$\alpha$.

WHI J2259+58: Large Galactic nebula, probably emission, though we didn't get an
H$\alpha$ image.

WHI J2309+44: Turbulent Galactic nebulosity.

WHI J2312+25: Arrow-shaped bit of Galactic nebulosity.

WHI J2319+43: Galactic nebulosity.

WHI J2353+70: LSB spiral.

{\bf Southern Candidates:}

Five objects were not included in \citet{WHI02} because our analysis of follow-up
images did not appear to show anything at those positions.  However, reprocessing
of the original data and remeasuring allow us to list them:

WHI J0551-39: A large ring, possibly a very old PN.

WHI B0740-02:  The envelope of the cD galaxy in a cluster.

WHI J1019-23: A vaguely elongated patch of light.

WHI J1129-13: There is a patch of light which is not the
right shape for the flat-field residuals.  It's at rather high Galactic
latitude for nebulosity (though see the note on WHI J1050+64), so it could
conceivably be extragalactic.

WHI J1828-52: A large ring?

Conversely, attempts to measure the surface brightness of WHI B1249-33 result in no
significant signal.  There is probably something there, since an object was seen
on both the red and blue plates; but it is too faint to measure (a more extended
discussion of non-detections appears below).  It is, however, possible that we
were fooled by a pair of coordinated plate defects along with a flat-fielding 
error, and it is not certain that this object exists.

WHI B1751-07 was listed as a PN in \citet{WHI02}, but on re-examining the image it's 
pretty clearly a spiral galaxy.

\subsection{Selected Images}

While we have images of each of our objects, including them all would clearly make this
paper quite unwieldy.  Instead we show here a selection, both of representative examples
and of objects of special interest.

\begin{figure}
\plottwo{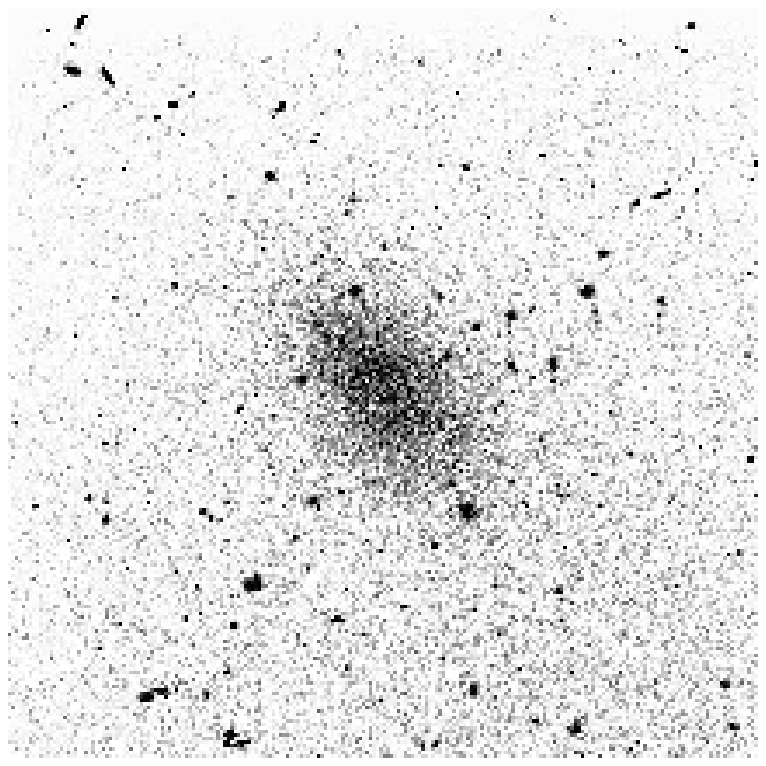}{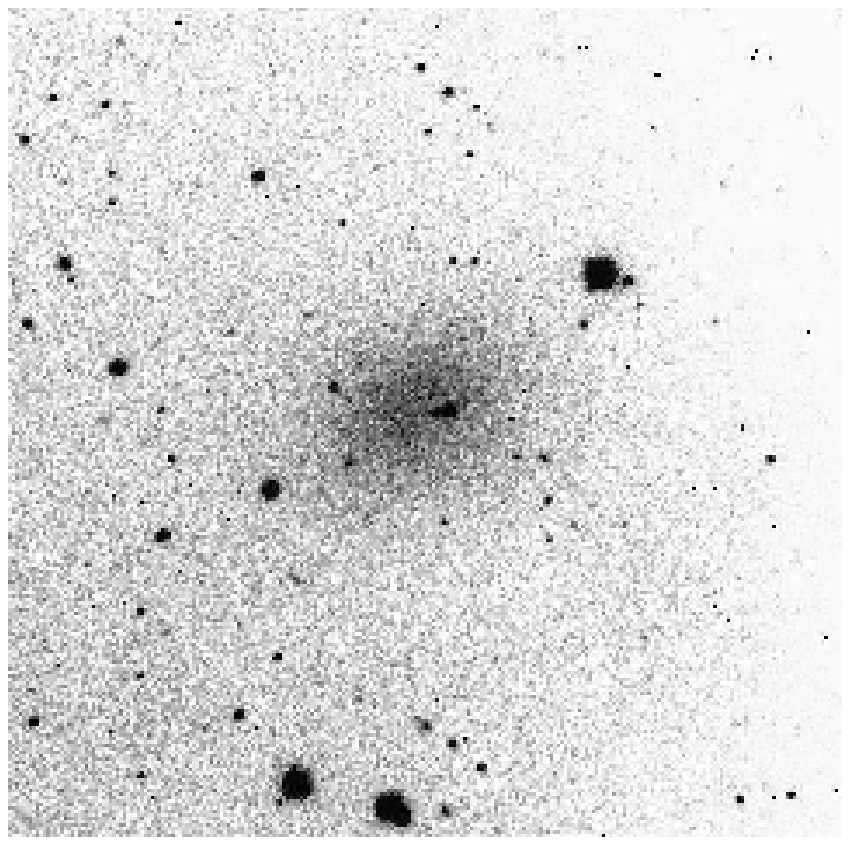}
\caption{Two of the distant (non-Local Group) galaxies found during our survey: on
the left, UGCA 275 (field about five arc minutes square); on the right, KKH01-59 (field
about four minutes square).  Each image is a 900s exposure in $R$.
In all images, here and subsequently, north is up and east to
the left, and all were taken with the KPNO 2.1m.}
\label{spheroidal}
\end{figure}

\begin{figure}
\plottwo{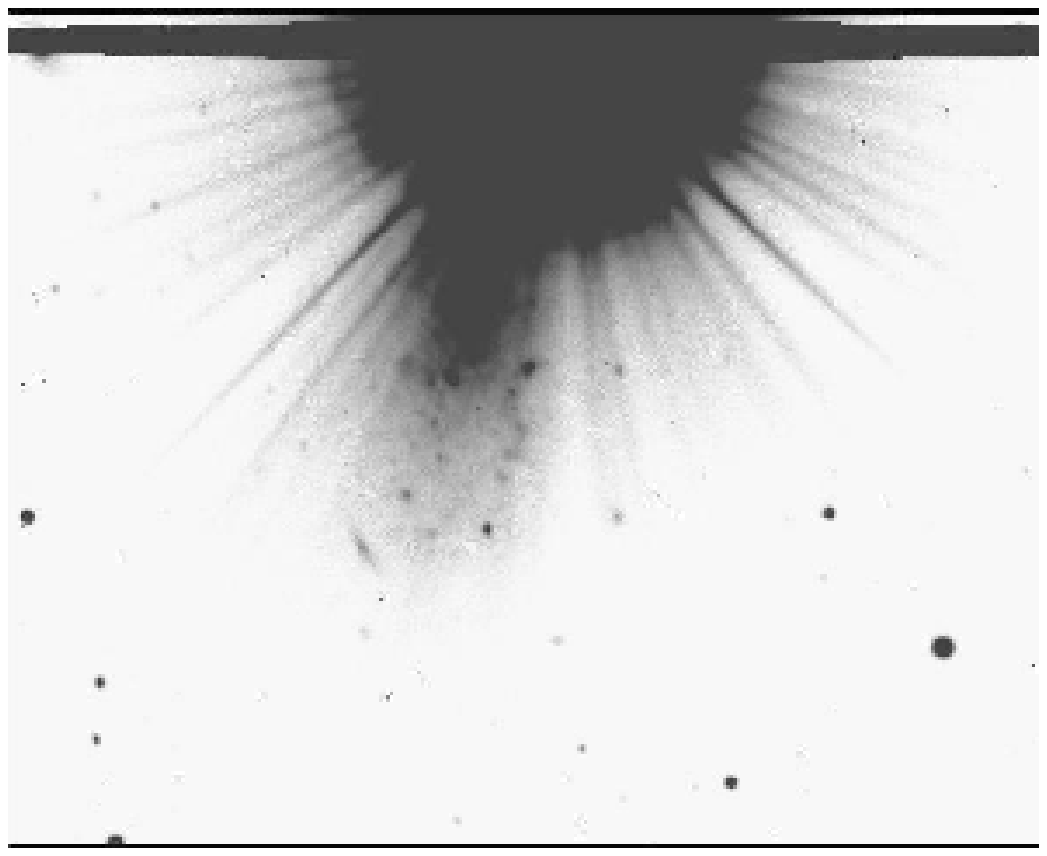}{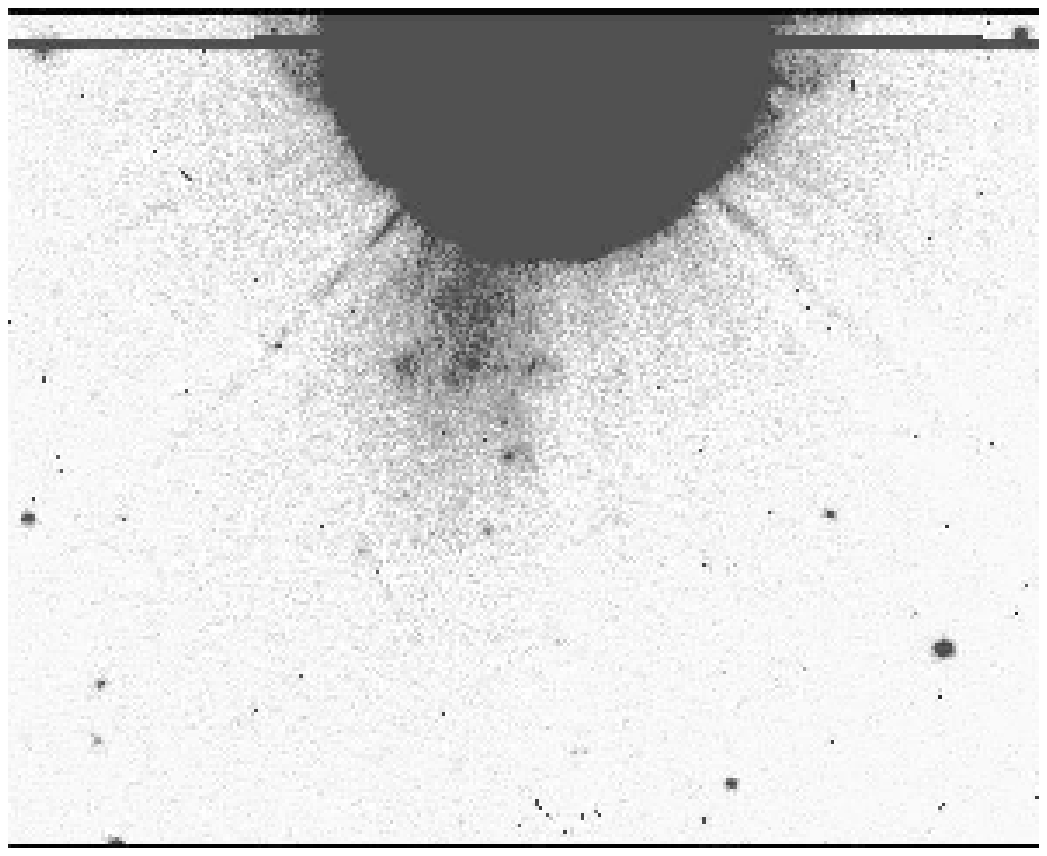}
\caption{An inconvenient bright star interferes with the study of PGC 39058.  Left,
the galaxy appears to be resolving in a 900s $R$ exposure; at right, some of the
discrete objects are shown to be HII regions in this 1200s H$\alpha$ exposure.
Both images are about four arc minutes high.}
\label{interference}
\end{figure}

\begin{figure}
\plottwo{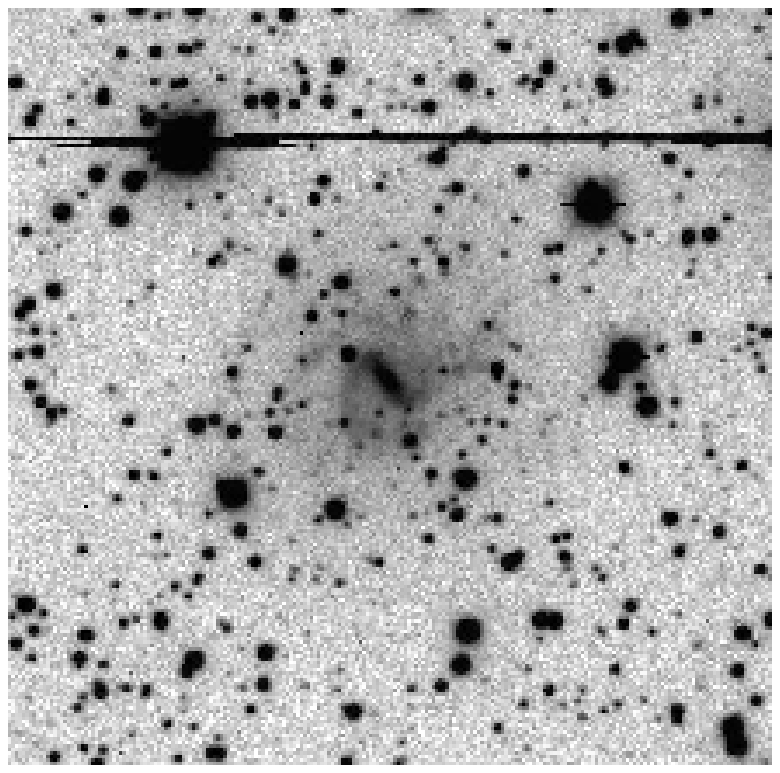}{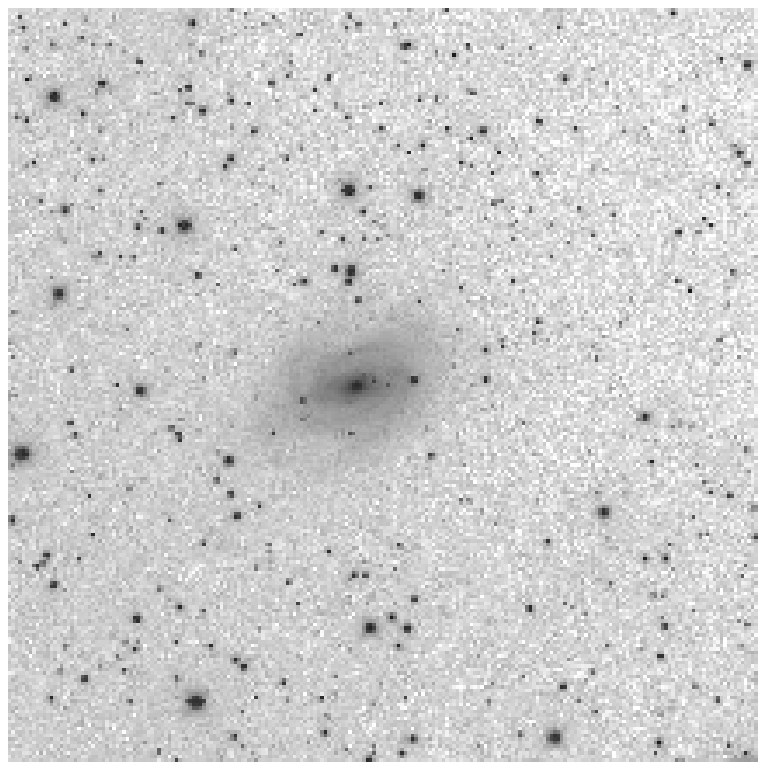}
\caption{Distant, face-on spiral galaxies; on the left, CGMW 5-5772
and on the right, ZOAG G134.31+06.24.  Both fields are about four arc minutes
square and both exposures were 900s in $R$.  On the survey plates they resemble
faint, diffuse dwarf galaxies (the spiral arms) with a superimposed star (the 
bright nucleus).}
\label{spiral}
\end{figure}

\begin{figure}
\plottwo{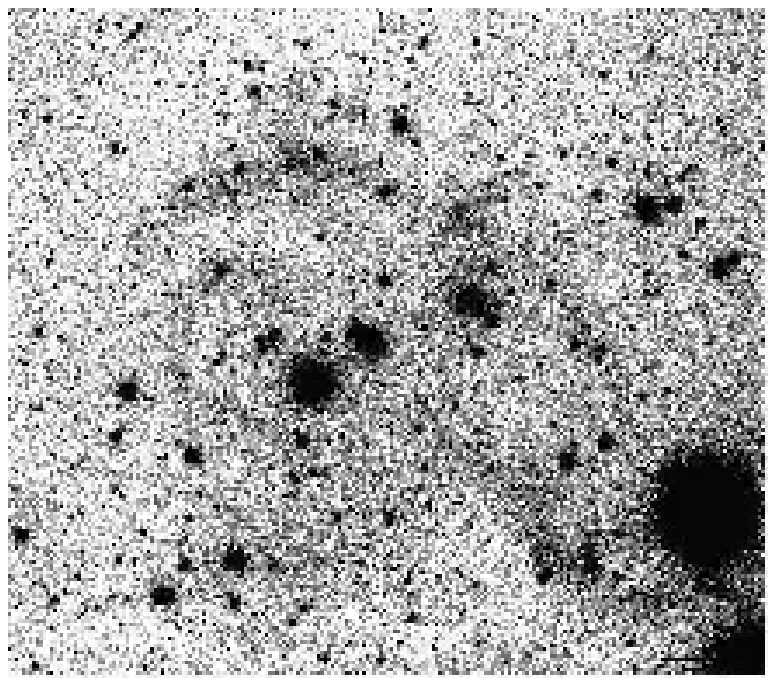}{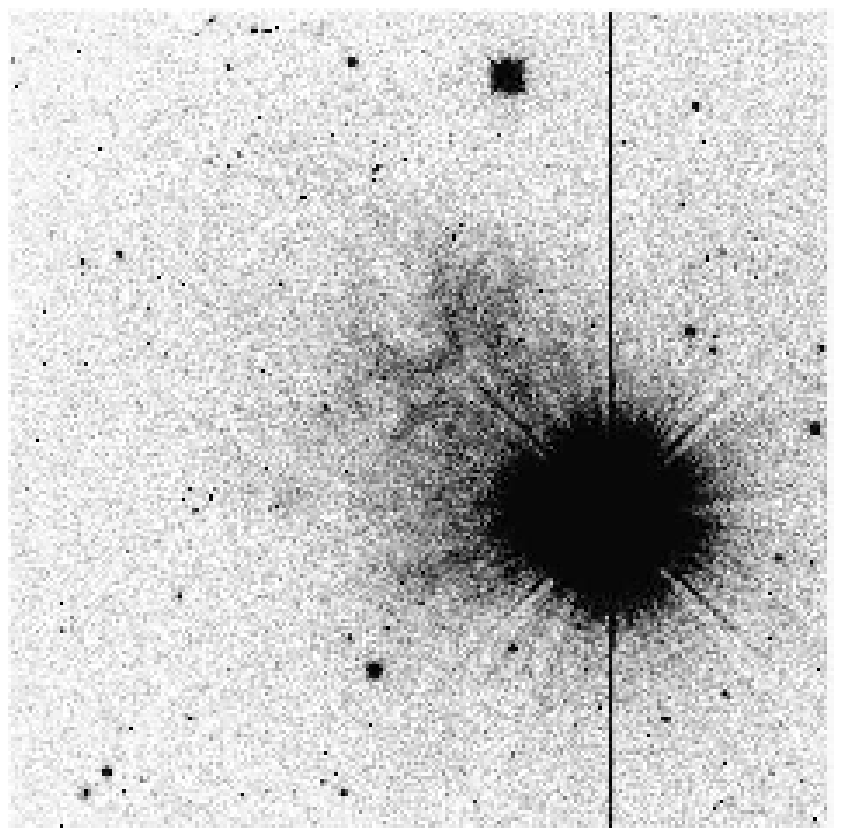}
\caption{Two examples of Galactic nebulosity from the list of candidates.
On the left, WHI J2004+64 is a section of Galactic cirrus, whose exact
center is clearly difficult to define (field about 9 x 10 arc minutes; 900s
exposure in $R$).  On the right, WHI J0156+13 is apparently involved with
the bright star HD 11861, though we cannot rule out a line-of-sight
coincidence (900s exposure in $V$, field about four minutes square).}
\end{figure}

\begin{figure}
\plottwo{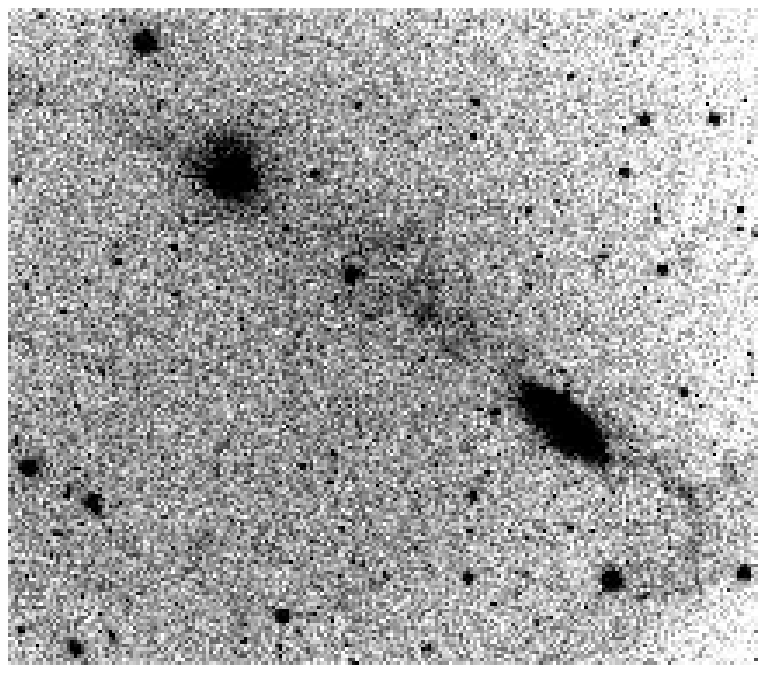}{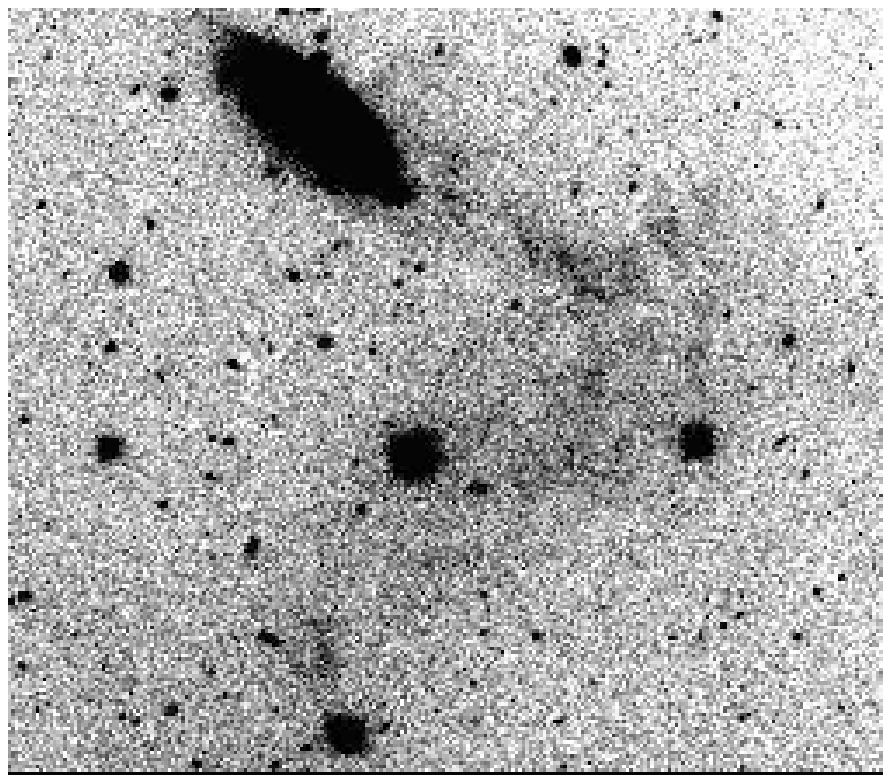}
\caption{Galactic nebulosity coincidentally aligned with the distant galaxy
UGC 5932.  Left, centered on the knot catalogued as BKK 7, which appears in
our tables as WHI J1050+64b; right, showing the knot (which also appears at
lower right in the first picture) WHI J1050+64a.  The left image is the
full frame of the KPNO 2.1m, slightly over 10 arc minutes on a side; 900s
exposure in $V$.  The right image is just under 5 arc minutes high,
900s exposure in $R$.}
\end{figure}

\begin{figure}
\plottwo{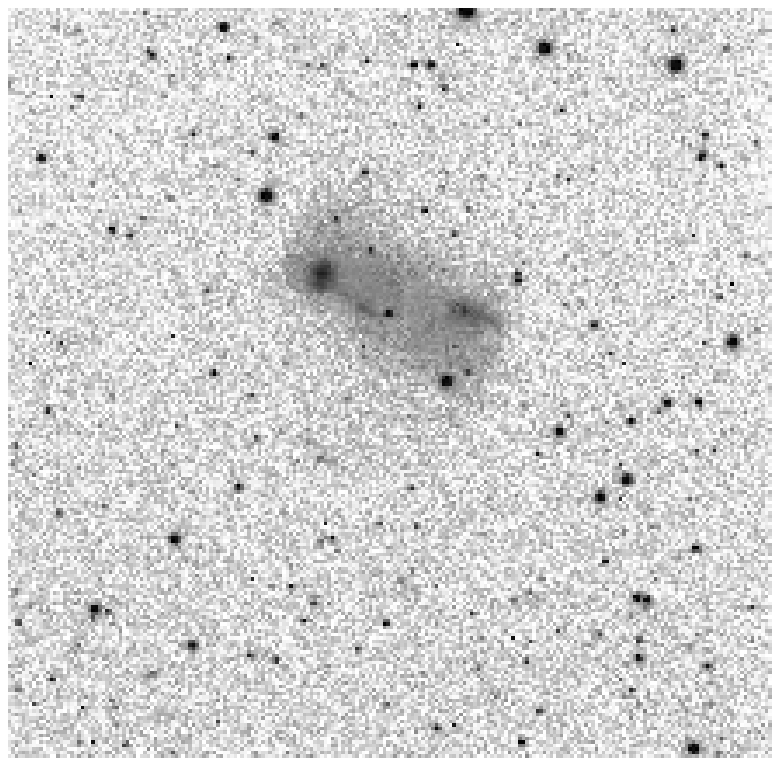}{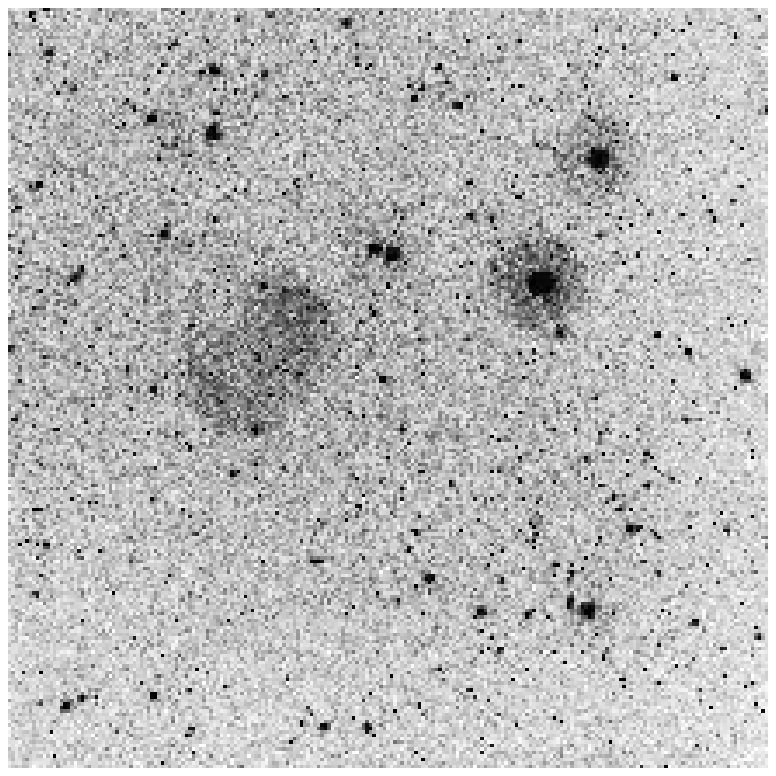}
\caption{Two planetary nebulae from the candidate list.  Left, ZOAG 139.32+04.85,
also known as KK98-26 and Camelopardalis C; 1200s exposure in H$\alpha$, about
five arc minutes high.  Right, WHI J 1919+44, also 1200s in H$\alpha$, field about
ten arc minutes high.}
\end{figure}

\begin{figure}
\plottwo{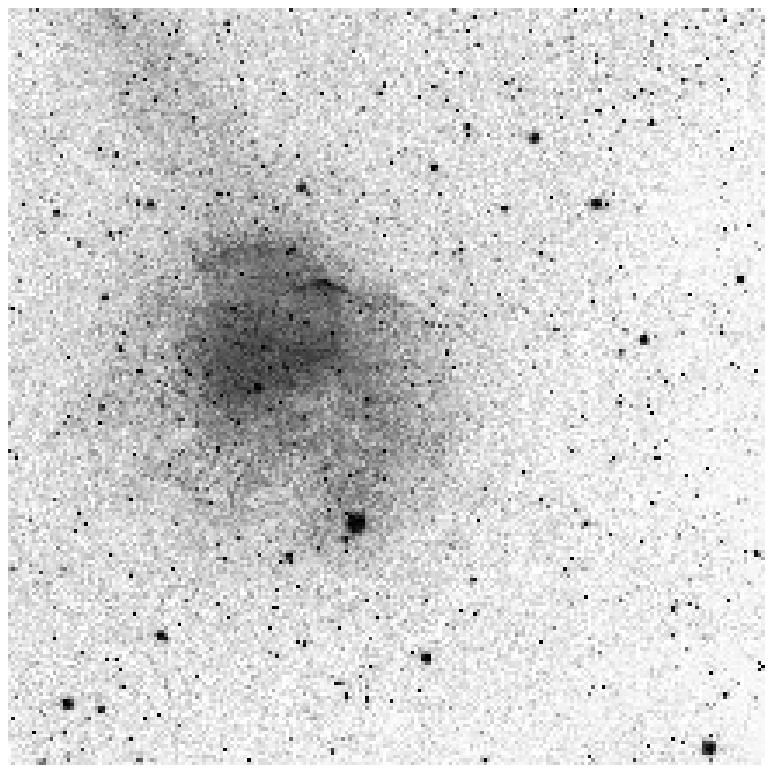}{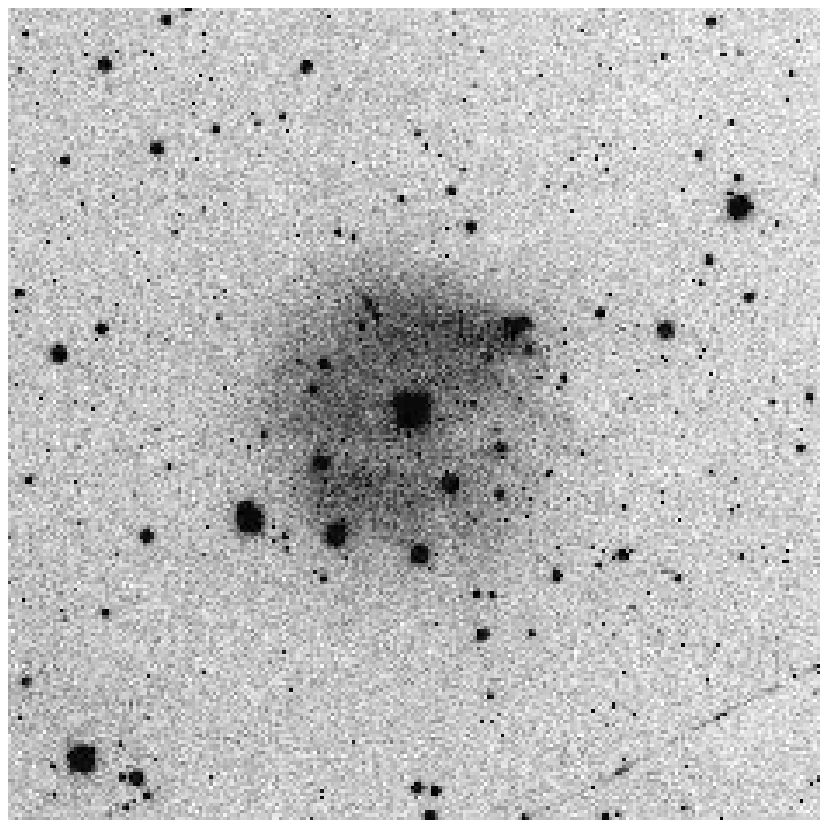}
\caption{Two more Galactic emission nebulae from the candidate list.  Left,
Sharpless 141 appeared in a catalogue of HII regions in 1959, but the fact
did not come to our attention before we had obtained this 1200s exposure
in H$\alpha$ (field about 10 arc minutes square).  On the right, WHI J0018+65
does not seem to have been catalogued previously; also 1200s in H$\alpha$, field
about five arc minutes square.}
\end{figure}

\clearpage

\section{Analysis and Discussion}

The data presented in the previous section contribute to several areas of dwarf
galaxy studies by presenting new examples and more definite measurements of known
objects.  (As a byproduct we have presented similar data on Galactic nebulae.)
We note that all four tables, taken together, constitute an all-sky
sample of objects selected under uniform morphological criteria, without regard to
(for example) redshift, Galactic latitude,
or membership in any particular group.  As such, it
has much potential use in comparing with samples selected in different ways.

However, there is more information to be derived from our work.  As the
only (to our knowledge) deep, all-sky survey for Local Group galaxies it
has the potential to set firm limits on the census of the Group.  To do
that we need to be able to say, as accurately and quantitatively as possible,
what we could {\em not} find.  We do this in two parts: first, by comparing
three different data sets, we estimate the reliability of our survey; that
is, we estimate what fraction of the objects visible on the plates and
meeting our selection criteria were actually detected.  Second, by examining
the measured surface brightness distribution of the follow-up observations
we estimate the sensitivity of our survey.  Strictly speaking the two aspects
are related, and routinely one derives a selection rate as a function of
surface brightness (and other parameters as appropriate).  Two features of
our survey lead us to separate calculations: first, the plate material forms
a fixed set of data, so that for instance by looking again one can determine
for certain whether an object is visible on it or not.  This means that one
can form a well-defined fraction of objects visible on the plates.  Second,
the peculiarities of a visual survey make it useful to separate the aspects,
as will be seen below.

\subsection{Reliability and Completeness of a Photographic-Visual Survey}

In an era of automatic algorithms working on electronically obtained and calibrated
data the idea of a visual survey on photographic material appears almost quixotic.
Certainly photographic data are less well controlled than CCD data, especially
if the former are in the form of an emulsion (instead of a digital scan).  
However, less well controlled is
not the same thing as uncontrolled.  \citet{MES90} found that the plate-to-plate
variation among 185 plates used in the APM galaxy survey (southern sky)
amounted to 0.178 mag rms.  We expect the consistency of the POSS-II plates to be 
similar, given that the various technologies involved were maintained or slightly
improved in the meantime.  The later northern survey is clearly of a lower quality
in the number of plate defects, aircraft trails and the like (reflecting both the
enormous difference in the level of air travel over the south 
and the limited time available to take duplicates);
but while this increased the time spent chasing chimeras it does not really change
the consistency of the photographic material.

The other part of the survey, detection of objects by the human eye (with its
very effective but rather complicated software), is a different operation from
automatic routines and needs to be appreciated in a different way.  The major
difference lies in the fact that a human observer does not apply search
criteria exactly and in a completely reproducible way.  This means that an 
object which clearly matches the criteria and is well within any detectability
threshold may be missed anyway.  Contrast this with an automatic function, which
will {\em always} find (say) a round object (ellipticity less than a given
number) standing ten sigma above the background.  In addition, there will be
a wide region at the edges of search criteria in which a visual survey may
or may not include objects.
The latter becomes especially important when dealing
with the (necessarily) vague morphological criterion of looking ``like known
dwarfs: diffuse and faint.''

\subsubsection{The 211-Plate Comparison}

For various reasons of timing, 211 fields of the POSS-II survey were examined in
1997 and again in 2000.  This is certainly enough time (and enough sky area) to
remove any memory of the first examination by the time of the second, so they
may be considered to be independent looks at the same data.  

As an initial comparison we look at the estimate of general Galactic nebulosity.  In terms
of plate grading for Milky Way interference, the numbers were \\
1997: good 124, troublesome 42, poor 13; \\
2000: good 132, troublesome 26, poor 21; \\
with a total of 32 fields changing grading by one column, up or down (none by two).
Given the subjective nature of the grading, this is a very stable performance.
More to the point, our adopted overall correction for Milky Way interference 
(one-half of the troublesome fields plus all of the poor fields) is identical.

We now turn to the survey itself.
To estimate the completeness of a survey, we assume a target population of 
some unknown number $N$ defined
by our set of observational criteria, each of whose members the survey has an 
average probability $p$ of detecting and recording.  These are the objects
fitting our morphological criteria and visible on the plates.
The number actually detected at one pass will be
$n_1 = p_1N$; since neither quantity on the right is known, we cannot 
calculate the other.
On a second pass (or with a different survey), similarly, $n_2 = p_2 N$ will 
be detected, allowing for the fact that the probability may change; and of 
course the particular set of objects will differ in general, with an overlap of 
$n_3$.  If we assume the first set of detections $n_1$ is a random sample of the 
parent population, the probability of detection by the second survey is just 
$p_2 = n_3/n_1$.  Working backwards we can now calculate $p_1$ and $N$.  
If the two probabilities are significantly different, we
should start looking for systematic differences between the detected objects.

Looking at the number of candidate objects in this way, things are less encouraging
than with the plate grading.  36 were
recorded both times; 20 only in 1997; and 31 only in 2000.  Taken at face value,
it seems that an object matching our morphological criteria has slightly over half
a chance, certainly not as much as two-thirds, of even being seen.

But on a third examination of these fields we found that, of the 1997-only
objects, 17 were really too bright and small to fit the criteria; and of the 
2000-only objects, 26 were actually too much like Galactic nebulosity to be worth
recording.  What happens is this: in each examination we were anxious not to miss
any true Local Group dwarfs, and so included many doubtful objects.  Between
the two examinations there was a shift in which doubtful objects we were
inclined to include.  A second look at each candidate showed that many were
clearly not of interest, and that we were choosing good candidates much more
reliably.  If we include only good objects we come up with a one-pass
reliability approaching 90\%.
This is probably optimistic, however, since it only compares one set of eyes with
itself.  To get a better comparison we need another set of eyes.

\subsubsection{Comparison with the Local Volume Survey}

We are aware of only one other survey comparable to ours, that is, which examined
photographic material over the entire sky (or a large fraction of it) in seach of
faint objects with the morphology of dwarf galaxies.  As reported by \citet{KKH01},
their results along with those of \citet{KK98}; \citet{KKR99}; \citet{KKS00}; and
\citet{KK00} covered 97\% of the sky looking for dwarf galaxies in the Local Volume
(out to a few megaparsecs), using film copies of the ESO/SRC and POSS-II plates.
While their criteria were slightly different (for example, they included objects
down to half an arc minute in size, which of course reasonable if one is interested
in things farther away than the Local Group) they should have included 
our criteria as a subset.

Going through the lists of objects as published, taking only those larger than
one minute along at least one axis and appearing on both the blue and red films,
the situation again appears rather distressing.  There is an overlap of 46 objects;
set against our candidate list of 194, this gives the Local Volume group
a chance of less than one-quarter of recording an object meeting our criteria.
Even if we confine ourselves to objects identified as
extragalactic the Local Volume probability of detection only reaches about 36\%.
There are similarly something like 200 objects in their list
which appear from the tables to match our criteria, but do not appear in ours,
giving us a chance of less than one in five.  
It seems hardly likely that even a cursory
visual survey would miss {\em most} of the objects it was looking for.

Again we turned our attention to the non-overlap region.
Visually examining the objects in the Local Volume list which nominally 
match our criteria, almost all are too small
and/or too bright to look like Local Group dwarfs to us.  When faced with an object
close to a minute of arc in size on the survey field, we would tend to say it is
smaller than the Local Volume group would report it; and there are enough objects
near the borderline to make a big difference in the numbers.  The fact that all of
these are very diffuse objects, whose size is difficult to estimate anyway, only
complicates matters.  If we count up those objects which we did judge to meet
our criteria, and include some borderline cases, we find 14 which
we missed.  This gives our group an estimated reliability 
(conservatively) of about 77\%.  (A similarly
corrected number for the Local Volume survey is not available without asking them
to review our candidate objects, clearly an unreasonable request.)

We wish to make clear what is happening in this process: the human eye and
judgement, when applying a set of morphological criteria corresponding to known
Local Group dwarf galaxies to images, generates a fairly reliable (77\% or maybe 
better) list of those that fit.  It will also (probably to avoid missing anything
important) throw up a large number of doubtful cases.  These can be rejected
fairly easily by a second, independent examination.  We may make an analogy 
with the classification of galaxies by eye, based on morphology: the majority
will be assigned the same, or very similar, types by different workers; but
there is a population whose assigment will vary widely between surveys.

\subsubsection{Comparison with Known Local Group Galaxies}

Seeking a way to check our completeness without dealing with the diffuse edges of
morphology space, we turn to a sample which is very definitely made up of objects
morphologically like Local Group dwarf galaxies: known Local Group dwarfs. We compared
our Local Group detections in Tables \ref{northcand} through \ref{southnon} with
known and suspected Local Group galaxies (defining them observationally, leaving aside
such questions as whether the Sextans A/B/NGC 3109 subgroup is actually bound to the
Local Group).  We found that Pegasus (DDO 210), Leo A (DDO 69), and Phoenix, as
well as anything brighter, were too bright for us to record.  Sextans is not visible
on the survey plate, and so is outside our morphological criteria; similarly for
the recent detections Andromeda IX and Ursa Major.  Ursa Minor is visible, as are
Andromeda V and Cassiopeia, but these were not recorded.  We could argue that 
Cassiopeia looks rather like a Galactic nebula, and is found on a ``troublesome'' 
plate, so could be counted under the allowance for Milky Way obstruction 
(see below); but in the interests of a conservative estimate we include it.

Seventeen detections out of a total of twenty gives a detection probability of 85\%,
about midway between the previous figures of 77\% and 90\%.  Given the small
numbers involved and the uncertainty in applying the morphological criteria, 
this is a remarkable level of agreement between three independent data sets.

\subsubsection{The Number of Missing Galaxies}

Using the lowest of the above figures along with the total of 20 for Local Group
dwarf galaxies of the targeted type, we come up with an estimate of 26 for the total
population.  That implies that (statistically) six objects with
similar properties are still waiting to be found.
However, if we include the Local Volume survey and grant them the same probability
of detection as ours, the joint probability of detection goes up to 95\%, implying
just one remaining undetected Local Group dwarf within our detection
criteria.  (This assumes that all the Local
Volume objects have been followed up in enough detail to demonstrate that they are
Local Group dwarfs, or not.  From the sample of 14 objects referred to in the
previous subsection, nine can be ruled out from online data alone.  We are looking
into the situation on the remaining five.)  In practice we believe this to be a low
estimate, since it does not take into account all the other groups which have
looked at the sky in various ways.

We conclude that it is unlikely, but would not be surprising, if there remained one 
Local Group dwarf galaxy sufficiently large and bright to appear on 
the survey plates, and not hidden by the Milky Way,
which has not yet been found.  Two would be surprising; a larger number is extremely
unlikely.

Again we wish to emphasize the differences between a visual survey and one using
automatic algorithms.  First, there will be objects missed regardless of how easy
they may be to find; second, an effort not to miss something important will generate
many doubtful candidates that on review should be discarded.  In spite of the
inherant fuzziness of such a survey its repeatability (and hence completeness)
can be characterized with some confidence, as the agreement of our three estimates
shows.

\subsubsection{Milky Way Interference}

The important caveat to the previous section's conclusions is that the Milky Way
covers a large part of the sky, and none of the objects we are looking for can be
seen through Galactic interference.  To estimate the completeness of the Local
Group census we must come up with an estimate of Galactic sky coverage.  A simple
expression in Galactic latitude is rather too crude (especially in light of the
significant cirrus found above $b = 45\arcdeg$).  A sophisticated treatment
involving optical extinction integrated over the sky would require 
effort disproportionate to the result we seek (and to the 
uncertainties in the other assumptions we must make).  Instead, we have, as noted
above, separated all survey fields into three grades.  We assume that a Local Group
dwarf would be seen over all of a ``good'' plate; over half of a ``troublesome''
plate; and not at all on a ``poor'' plate.  From this procedure we estimate that
72.5\% of the sky is clear of Milky Way obstruction.  (For comparison, we note
that \citet{WG04} estimate 67\% of the sky as being free from Galactic
interference using an entirely different method, agreement well within our
expected uncertainty.)

Converting this to an estimate of the number of obscured galaxies involves other
assumptions, particularly about the luminosity function in the Local Group.  
Certainly the brightest galaxies would be seen even through much of the Milky
Way, and gas-rich systems detected through (for example) HI observations, and so
for them the 72.5\% is too low.
Taking our 20 faint galaxies and adding, say, ten more from the known Local
Group population (which would be obscured,
but were too bright to meet our search criteria); and assuming an isotropic
distribution, we come up with an estimated eleven galaxies in the obscured zone.
If we take the extreme position that all of the ``troublesome'' plates are
completely obscured, we have only 58\% of the sky clear, and an estimated 22
hidden galaxies.  This is an unlikely number (we {\em did} find faint extragalactic
objects on ``troublesome'' plates), best taken as an extreme upper
limit.  

The estimate changes if the distribution of galaxies is not isotropic
on the sky, of which there is some evidence, but it has not been clear 
whether the
distribution is more concentrated toward the Galactic Plane, as found (for
instance) by \citet{B05};
or less, in accordance with the venerable Holmberg effect, most
recently found in the results of \citet{SL04}.  The latter found a preference
for higher-latitude ($> 30\arcdeg$) positions over lower ones in the ratio
of 0.8; the former's results are a bit harder to put in the form we are 
looking for, but amount to something less than 1:1.5.  The work of
\citet{Yv06} confirmed the result of \citet{B05}, showing an excess of
satellite galaxies along the major axis reaching 20\% and a similar deficit
along the minor axis; they also explain the discrepancy with 
\citet{SL04} as due to an ambiguity in defining the major axis
angle on the sky, concluding that all three studies are in fact
consistent.  But they also found that, for blue central galaxies (like the
Milky Way), the effect disappears.  In any case, the correction implied
for the number of satellite galaxies hidden behind the Galactic Plane is
no larger than the uncertainty in the number of satellites in the sky area lost
due to Galactic interference.

\subsection{Sensitivity of the Visual Survey}

We may get a first estimate of the faintness limit of our survey by comparing those
Local Group dwarfs found and missed.  According to the compilation of \citet{M98},
Draco (which we found) has $\Sigma_V = 25.3 \pm 0.5$; Ursa Minor (which is visible on the plate,
but we did not notice) has $\Sigma_V = 25.5 \pm 0.5$; Sextans, which is not visible,
has $\Sigma_V = 26.2 \pm 0.5$.  This indicates our limit is about $\Sigma_V = 25.5$.  For
galaxies such as Andromeda IX \citep{ZK04}, with $\Sigma_V \sim 26.8$, we had no chance.
And while a surface brightness can be calculated for such objects as the Canis
Major \citep{M04} and Sagittarius \citep{IGI95} galaxies, such a figure is not
observationally significant, since the dwarf galaxy's stars are lost among those of
the Milky Way.  (We are inclined to place them among the galaxies lost to Milky Way
interference in the calculations of the previous section.)

For later use we would also like to place a bright limit on our survey.  Again
comparing our list to that of \citet{M98}, we find that objects at $\Sigma_V \sim 23$
or brighter were always excluded; a few tenths of a magnitude fainter, no object
was excluded for being too bright. The matter is made uncertain by the lack of
measurements on some galaxies, the large uncertainties on others, and the inclusion
in our list of brighter objects that appeared to be resolving or on the verge of
resolution, but a figure of about 23.3 is probably our best estimate.

More important is a better estimate of the faint limit, especially given the 
large uncertainties in
the numbers quoted.  At the very least, a confirmation of the general figure of 25.5
by a different method is desirable.  However,
in trying to estimate the limits of sensitivity of our survey, as well as in trying
to determine the nature of each candidate object, we run up against the nature of
the follow-up observing.  It was designed to answer one question for each object:
is it a Local Group dwarf galaxy?  There was simply not sufficient telescope time
for the multiband and multimode observations required to say for sure {\em what}
each object is.  Neither were all data taken under the photometric conditions which
would allow a precise measurement of brightness; in fact true photometric conditions
were rare.  So we are faced with pressing into service data taken for another purpose.

In addition, measuring surface brightness, especially of faint and diffuse objects,
is difficult both in definition and in practice.  For a large and complicated Galactic
nebula, for instance, where is the center?  If that cannot be defined, then neither can
one define a central surface brightness.  If the object is not of some regular form,
then the apparatus developed for comparing elliptical galaxies (say) is not of much
use.

In spite of all this, it is possible to obtain
some measurement of how faint an object the eye-plate combination
could detect.  To this end, the flux within a section of each candidate was measured
(using the IRAF routine polyphot), an area roughly
one arc minute in diameter containing the brightest parts.  Stars were excluded as
much as possible (which was difficult near the Galactic Plane), as were the bright
nuclei of face-on galaxies (which appear starlike on the plates).  The intention was
to measure as closely as possible what the eye responded to in the survey field.
Then two (sometimes more) sections of blank sky were likewise measured, trying to
bracket the object sideways to the remaining flat-field gradients; these were averaged
and subtracted from the object flux.  The counts per pixel were transformed into
magnitudes per square arc second using photometric solutions derived from Landolt
standards taken the same night.  An uncertainty was derived using both the variation in
the sky readings and the uncertainty in the photometric solution.  The sky subtraction
dominated most measurements, even on non-photometric nights.  In several cases the
same object was measured on different runs; the results agree to within our stated
errors, though it was clear that the uncertainties are not overstated.

The measurements were all done in $R$, because only in that band do we have data on
all objects.  This is a result of looking for the tip of the Red Giant Branch, which
is easiest to detect there.  Unfortunately, most surface brightness estimates and
measurements are given in $V$; this must be borne in mind when making comparisons.

\begin{figure}
\plotone{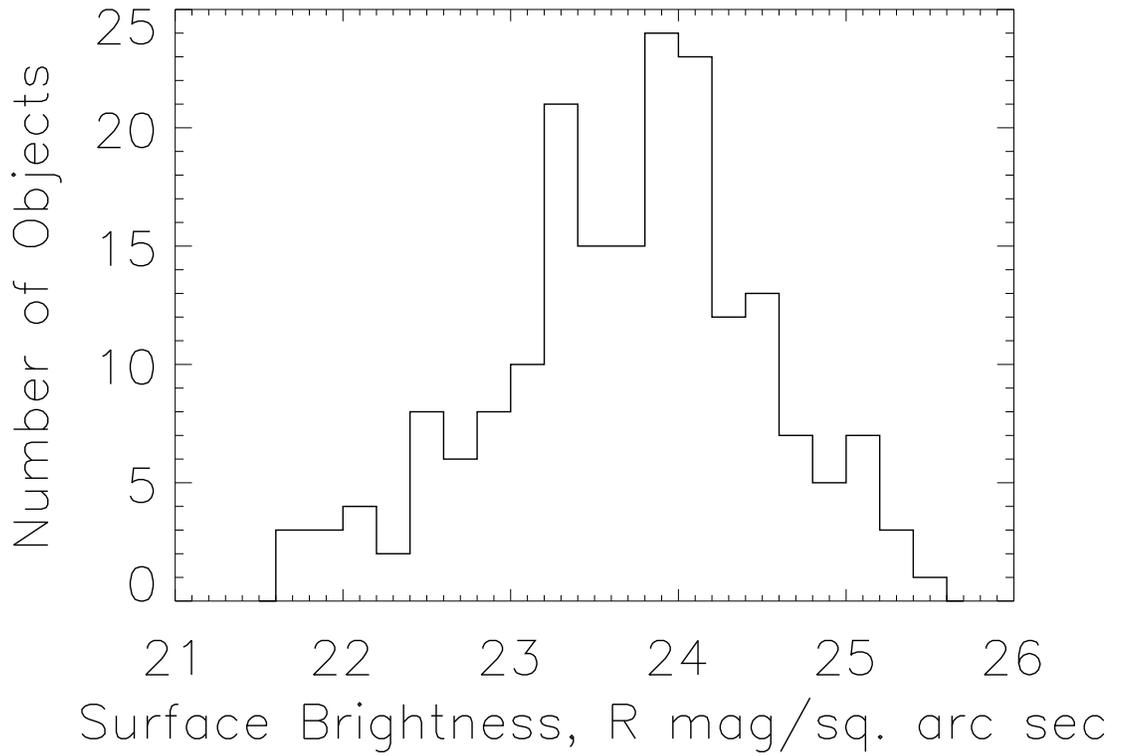}
\caption{The distribution of ``central'' surface brightness of all candidates, as
measured in the $R$ band from follow-up images.  Bins are 0.2 mag wide, which is
roughly half of the average uncertainty.  While the lack of objects fainter than
about 25 magnitudes per square arc second is almost certainly due mostly to the limits 
of sensitivity somewhere in the survey process, 
to derive a quantitative limit requires some model for the underlying
surface brightness distribution.}
\label{total}
\end{figure}

\begin{figure}
\plotone{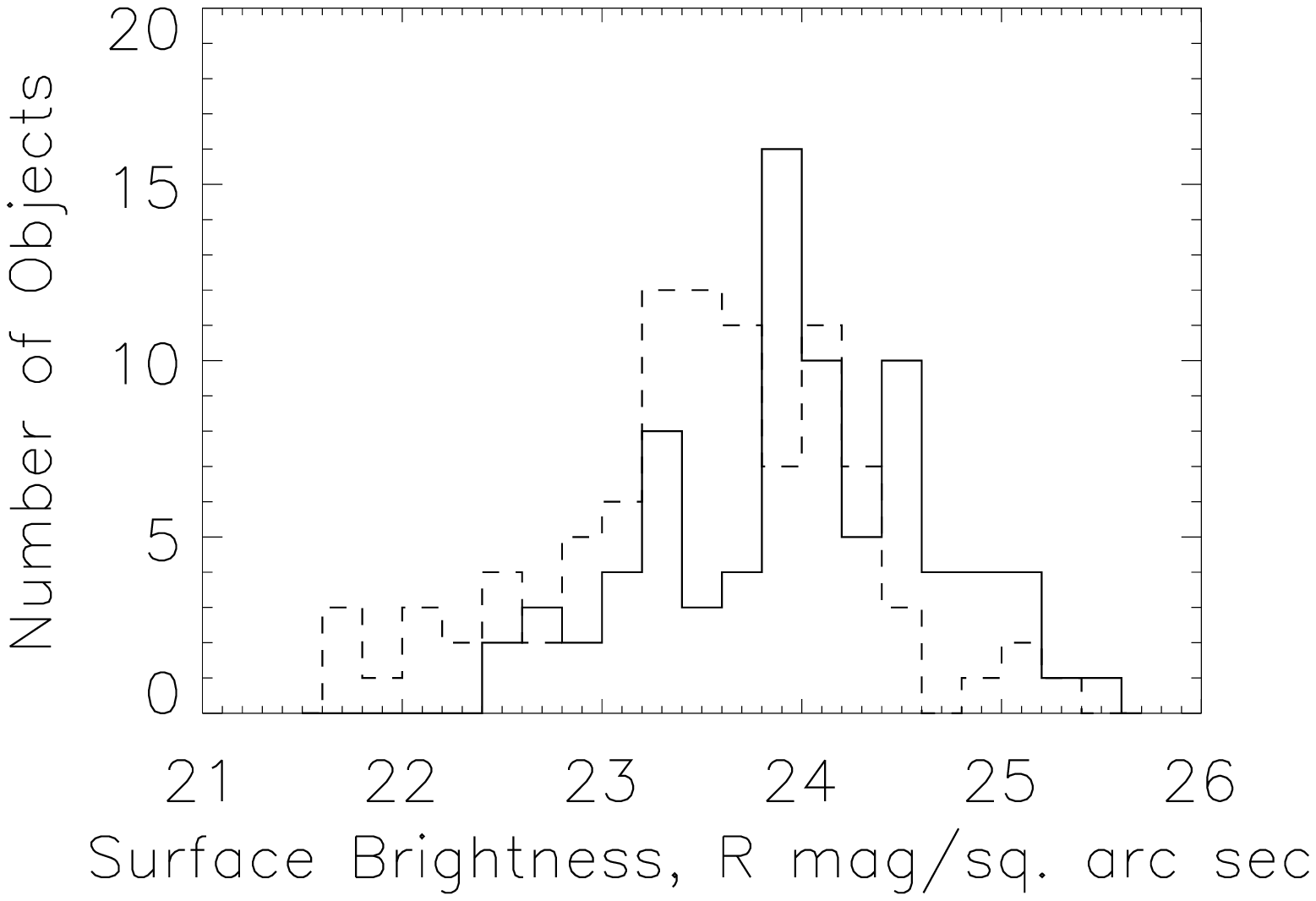}
\caption{The distribution of surface brightness for Galactic candidate objects (solid line) and
extragalactic ones (dashed).  Objects whose nature was uncertain are not included.
The Galactic distribution is clearly shifted faintward of the extragalactic, indicating that
a significant part (at least) of the fall-off in the latter beyond 24 magnitudes per square
arc second is real and not a selection effect.  A similar decline has been found in
large-scale galaxy surveys.}
\label{gal_xgal}
\end{figure}

The combined histogram for all candidate objects is shown in Figure \ref{total}.  Clearly we are
reliably detecting things out to about 24 mag arc sec$^{-2}$.  The reduction in number
per brightness bin beyond that point could be interpreted as the onset of incompleteness,
though if we're seeing a significant number fainter than 25 it's hard to understand why
we would be missing many a full magnitude brighter.  In order to transform the graph
into something quantitative, we need some model for the underlying distribution of
surface brightness.  (If we were dealing with {\em total} brightness we could apply
geometric arguments, but for the distances we are dealing with, surface brightness is
constant.)  While these do exist for galaxies, modelling such a thing for Galactic
nebulosity is a daunting thought.  For this reason we plot the two classes of object
separately in Figure \ref{gal_xgal}.

Here the peak and faint-end falloff of the Galactic nebulosities are clearly fainter than
the corresponding features for extragalactic objects.  This indicates that much of the
decrease between, say, 24 and 25 magnitudes is real, a feature of the underlying population,
and not due to a declining sensitivity in our survey at these levels.  Before we look
at the extragalactic data in more detail, however, there is an observational matter to
check into.

\begin{figure}
\plotone{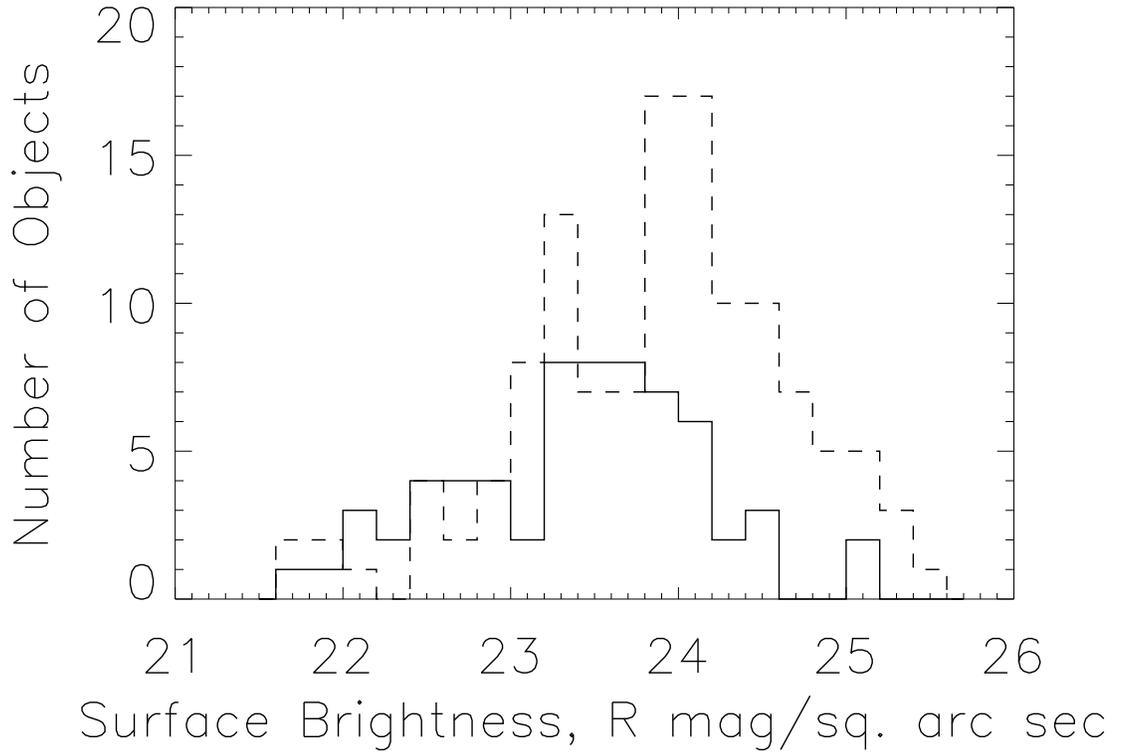}
\caption{The distribution of surface brightness for Local Group dwarf candidates, divided
into hemispheres.  Northern objects are shown dashed, southern ones solid.  No differences
in survey material or in the collection of objects themselves explain the much higher
relative numbers of faint northern objects; instead, the difference is traced to the
differences in follow-up telescopes.}
\label{northsouth}
\end{figure}

While processing the data we noticed an asymmetry between the northern and southern objects
in surface brightness, and so we plot them separately in Figure \ref{northsouth}.  Here is
a very clear trend: the northern histogram tends a full magnitude fainter at the faint end.
What could possibly cause such an effect?  The ESO/SRC and POSS-II surveys used essentially
the same emulsions and similar exposure times.  The Schmidt cameras involved were not
identical, but none of the differences should amount to anything like a full magnitude
(and most probably the brightness of the night sky was the most important limiting factor
in any case).  It is highly unlikely that there is such a marked relative lack of the very faintest
objects in the southern hemisphere of a few arc minutes in size.

But consider the follow-up telescopes themselves.  The CTIO 1.5m was operated at f/13.5, while
the KPNO 2.1m was used at f/7.5.  A faster system will record fainter diffuse objects in a given
span of time; even taking into account the shorter exposure times used in the north, the 2.1m
should theoretically have an advantage of a factor of 2.4, or about one magnitude, 
exactly as appears here.  (The Isaac Newton Telescope, used for a few of the northern
objects, has characteristics similar to the KPNO 2.1m).
The faint limit of the histograms here presented is due to
the follow-up observations, which did not reach as faint (in surface brightness)
as the original plate material.
This also accounts for the twelve objects found on the plates but not seen in the follow-up
images: they are diffuse, and too faint for the relatively slow telescopes to record.  (They
are not Local Group dwarfs, or we would have seen Red Giant stars in them.) 

Our histograms are thus a combination of (i) the original distribution of galaxies by
surface brightness, $F(\sigma)$; (ii) Galactic extinction; (iii) the selection function
of the eye-plate combination, $f(\sigma)$; and the selection function of the
follow-up telescopes, $S_N(\sigma)$ and $S_S(\sigma)$.  In addition, as we get to
fainter levels the number of objects diminishes greatly, adding to the uncertainty
of the statistics.  The mixture looks unpromising,
but with some plausible assumptions we can make a bit more progress.

\begin{figure}
\plotone{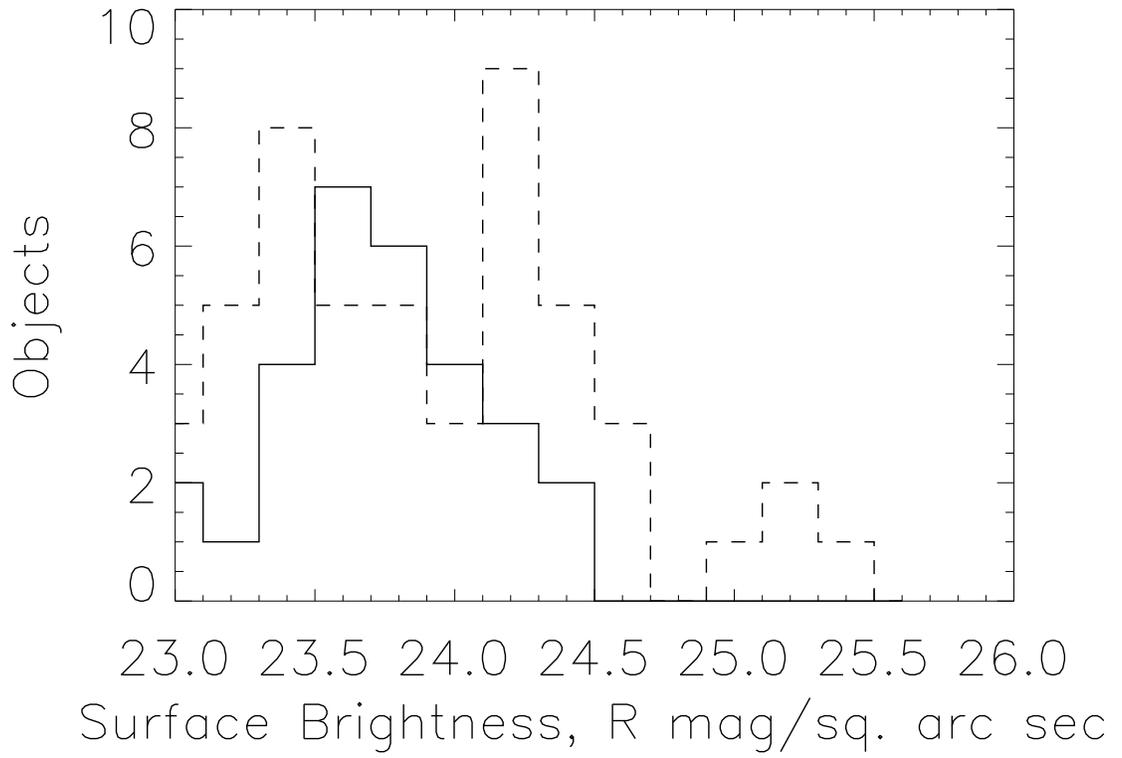}
\caption{The distribution of surface brightness for extragalactic
Local Group dwarf candidates, divided into hemispheres (south solid, north dashed).  
The data shown here form the basis for deriving the selection
function for the follow-up telescopes, as detailed in the text.}
\label{nsraw}
\end{figure}

We express the observed number of objects per bin in the northern sample as the
product of some total number, the original distribution as modified by Galactic
extinction and the plate-eye selection function, and the telescope selection function:
\begin{equation}
n_N = N_N F'(\sigma) S_N(\sigma)
\end{equation}
with a similar expression for the south.  Now we assume that $F'(\sigma)$ is the same
in the north and the south, and that the selection function for the 2.1m has the
same form as that for the 1.5m, only displaced by a certain amount.  (This displacement
could in principle be determined from the data, but not very reliably given the
small numbers involved, and in any case the inaccuracy of the original measurements
argues against trying to be too precise.  We will take it to be one magnitude.)  We
may now write an iterative expression for the telescope selection functions:
\begin{equation}
S_S(\sigma) = \frac{N_N}{N_S} \frac{n_S(\sigma)}{n_N(\sigma)} S_S(\sigma -1).
\end{equation}
We probably do not know the ratio of total numbers well at this faint end of the
function, so we will leave them as a normalization constant to be determined.
Taking only extragalactic objects, their histograms are shown in Figure (\ref{nsraw}).

We are clearly going to have trouble with the small number of objects per bin.  We
will take two measures to deal with that: first, smooth the data with a three-bin
running top hat (either before or after carrying out the division); second,
demand that the selection function be flat at brighter magnitudes (as long as
an object is bright enough, it should be seen) and monotonically declining
thereafter.

The ratio of southern to northern objects fluctuates in the bins from 23.0 to 
about 23.6 magnitudes per square arc second, depending
on the smoothing; partly this is due to small-number statistics, but possibly
also due to a somewhat ragged bright limit on our candidates.  At 23.8 to 24.0
it steadies, and declines thereafter.  We take this to indicate that there is no
brightness-dependent selection effect due to the 1.5m up to 24.0, and so none
for the 2.1m up to 25.0.  The next step is to apply our derived function to a set
of observed data and reconstruct, as far as we can, the histogram due to the original
brightness distribution and the eye-and-plate selection function.

As noted by the referee, our measured surface brightnesses include Galactic
extinction.  This would not be important for our purposes if it affected all
objects impartially.  Unfortunately, we find (using extinction measures from
\citet{SF98} via NED) that the more heavily extincted objects tend to be
brighter.  This makes sense: a faint object will be harder to find in a region
of high extinction because of confusion with bright Galactic nebulosity.  We
could in principle correct for this effect statistically, but the small number
of objects we have would make such a correction very uncertain.  Instead, we
simply omit those objects whose extinction is larger than the uncertainty
in their measured surface brightness.  This is equivalent to confining our
analysis to areas of the sky well clear of the Milky Way.

\begin{figure}
\plotone{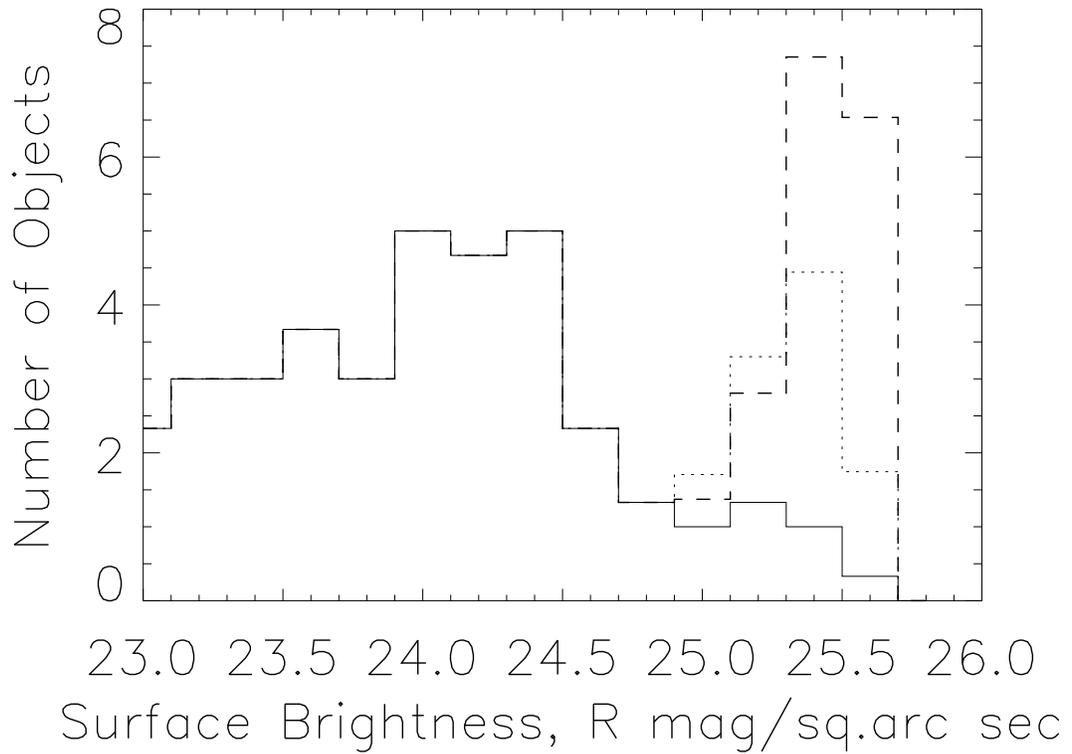}
\caption{The surface-brightness distribution of northern 
extragalactic dwarf candidates, smoothed with a three-bin running top hat
function.  The dashed and dotted lines show two reconstructions of the faint
end, removing the selection effect of the follow-up telescope (one smoothed
before computing, the other smoothed afterward).  This is the convolution
of the original galaxy surface-brightness distribution with the selection
effect of the visual survey on photographic plates, up to the point (at
about 25.6 magnitudes per square arc second) at which the 2.1m would not
show anything.}
\label{tophat}
\end{figure}

We now take the extragalactic northern objects, minus those with high extinction,
and apply a three-bin running top hat smoothing.  At the 25.0 magnitudes per
square arc second bin and fainter we apply a correction for the 2.1m selection
function.  The result is shown in Figure (\ref{tophat}).

There is a shallow rise to a maximum between 24.0 and 24.5, a drop thereafter,
then a rise again to the limit of the follow-up observations.  
The reconstruction
of the faint end, as shown by the two methods involved, is uncertain.  But it
is clear that, since the 2.1m showed any objects at all, there must be several
detected by the eye-and-plate method.  It is in this tail, we suggest, that
the objects not detected by follow-up imaging lie.

To work out specifically
the selection function of the eye-plate combination, even
as far as the limitations of follow-up observations allow, requires a knowledge
of the actual (surface brightness) luminosity function.  Unfortunately,
it is simply not known for galaxies this faint.  If there is a sudden
great increase in the number of objects fainter than 24.5 or 25 magnitudes 
per square arc second in $R$ our sensitivity there could be low indeed.  But we
note that both \citet{BL05} and \citet{DL05} have the surface brightness
luminosity function actually falling at their faint limits.  Since we go
deeper than either study it is certainly not impossible for there to be
such a faint upturn, but it would require some fine-tuning to have it
cut in at just the least convenient point.

The quantitative shape of Figure (\ref{tophat}) should not be taken too
seriously.  As we have noted, our data are relatively inaccurate and sparse,
especially at the faint end.  However,
we believe our overall result is robust: if the actual surface brightness
distribution of galaxies is not rising rapidly faintward of 25, and
extrapolation of current information has it indeed falling, then the 
combination of photographic survey plates and visual examination retains
most of its sensitivity out to 25.5 magnitudes per square arc second in $R$.

\section{Conclusions and Implications}

From our survey we conclude that the current list of Local Group dwarf galaxies is essentially
complete for objects brighter
than some limit between 25 and 26 magnitudes per square arc second in $R$
and larger than about a minute of arc
over 72.5\% of the sky.  (In total luminosity, 26 mag sec$^{-2}$ over a one-minute
diameter translates to $M = -7.6$ at a
distance of 1 Mpc.  We do {\em not} claim an absolute magnitude limit on our survey,
however.)  There may remain one more (improbably two) to be found.
Concealed behind the Milky Way we estimate another dozen; due to the uncertainty about
the luminosity function and spatial distribution of concealed galaxies this number is
uncertain, but unlikely to be as high as twenty.

Our survey was not capable of detecting all known Local Group galaxies.  Those whose stars
must be picked out from the field individually, such as Sextans, Sagittarius and Andromeda IX,
were invisible to us.  On the other hand, we did see many resolving objects near the Milky Way.
The requirement is for some perceptible increase in surface brightness (possibly in the form
of a stellar overdensity) above the surrounding field.  We note in particular that compact
star clusters of the type recently reported by \citet{H05}, if there were any significant number
associated with the Milky Way, would have been seen and recorded.  In that way and in others
the satellite galaxy systems of Andromeda and the Milky Way must differ intrinsically, since
the observed differences cannot now be attributed to selection effects. 

For many years the discrepancy between n-body simulations, which predict hundreds of galaxies
in structures like the Local Group, and the much smaller number of known galaxies has been
noted.  The simulations of \citet{KK99}, for instance, predict 300 satellite galaxies within
1.5 Mpc of the center of a Local Group-like structure, an order-of magnitude 
disagreement ``unless a large fraction of the Local Group satellites has been missed.''  
If that caveat is understood to mean that observationally accessible galaxies (that is,
those detectable over a large fraction of the sky in observations available
at the time the statement was made) had been somehow overlooked, it can now be removed.

The ``missing satellite problem'' persists in more recent simulations \citep{MM05}.  A popular
suggestion to deal with it involves reionization, which would make the conversion of small
dark matter haloes into visible galaxies much more difficult.  \citet{BF02} found that such
a prescription allowed agreement between observed and simulated Local Group data for a large
range of luminosities, producing a ``large population of faint satellites. . . 
awaiting discovery.''  In their simulation, the number of satellite galaxies brighter than
22 magnitudes per square arc second (in $V$) roughly matches observation; when the limit
was lowered to 26, the number increased to 200.  No numbers for intermediate levels are
given, but certainly the number per brightness bin must increase as one goes fainter, something
we have shown does not happen.  At
any rate, we have reached past 25 in $R$; for the galaxies made of old stars such as
\citet{BF02} constructed this is most of the way to 26 in $V$.
Their ``large population'' is simply not there.
(Note also that the detailed star formation histories of Local Group dwarfs appears to be
inconsistent with supression by reionization \citep{GG04}). 

Another suggestion (both of these are found in \citet{KK99}, though not original there) is
that small dark matter haloes were never able to form a significant number of
stars, retaining their baryonic matter
in the form of atomic hydrogen, and that these have been detected as high-velocity clouds
(HVCs).  Aside from the fact that no stars or other optical signal have been identified
with a HVC (for instance, \citet{SB02}), it appears that the HVCs are too small and nearby
to be identified with the missing dark halos, since any corresponding system should be
visible in nearby groups and isn't there \citep{PB04}.

The implications of our survey limit go beyond simple numbers of galaxies.  \citet{WG04}
identify a possible incompleteness in the census of Milky Way satellites beyond about
110 kpc.  Within that distance the faintest known galaxies can be found through
identifying their individual stars.  Beyond, if the limit for detection of diffuse
objects is 24-25 magnitudes per square arc second (their source for this figure and
the applicable waveband are not given), some unknown fraction of satellites would
not be seen.  If this incompleteness exists, the radial distribution of Milky Way
satellites could be the same as that of M31 satellites (and simulations).  Now,
however, with a limit of somewhere beyond 25, we can say that this possible 
incompleteness is not there.  We were sensitive enough to detect all but a handful of
known Local Group dwarfs, and those are more or less evenly divided between M31 and
the Milky Way.  The radial distribution of Milky Way satellites {\em is} different
from that of M31 satellites and simulations.

As this paper was in the refereeing stage several new objects even smaller and fainter
than Andromeda IX were reported \citep{Z06a, B06, Z06b, Z06c}.  
It is too early yet to draw any conclusions about their
implications for the missing satellite problem.  
If they are indeed the first of many (say, 100 or more) galaxies {\em and} their
dark-matter masses are of the same order as those of much brighter satellites, the
problem is solved (or at least transformed into a star-formation puzzle).  If not, the
problem persists.

\acknowledgements

This work has benefited greatly from a wide variety of sources.  Partial funding support was
received from the Intitute of Astronomy of the University of Cambridge, the Physics
Department of the U. S. Naval Academy, the European Southern Observatory, and two 
American Astronomical Society Small Research Grants (which last funds originated with NASA).
G. K. T. H. asknowledges financial support from the Chilean FONDECYT grant 1990442.
Extensive use has been made of the NASA Extragalactic Database (NED), which is operated
by the Jet Propulsion Laboratory, California Institute of Technology, under contract
with the National Aeronautics and Space Administration; and of the SIMBAD database,
operated at Centre de Donn\'{e}es Astronomiques de Strasbourg, France.  The results are
based on observations made with the Isaac Newton Telescope, operated on the island of 
La Palma by the Isaac Newton Group in the Spanish Observatorio del Roque de los Muchachos 
of the Instituto de Astrof\'{\i}sica de Canarias; and at Cerro Tololo Inter-American
Observatory (CTIO) and Kitt Peak National Observatory (KPNO).  Both CTIO and KPNO are
operated by the Association of Universities for Research in Astronomy, Inc. (AURA), 
under a cooperative agreement with the National Science Foundation, as the National
Optical Astronomy Observatory (NOAO).

We would like to express our thanks for the help of John Pilkington and Robin Catchpole
(then) of the Royal Greenwich Observatory, and Sue Tritton of the Royal Observatory
Edinburgh, with the thousands of survey plates we have handled; for the dedicated help
of the night assistants at Cerro Tololo and Kitt Peak; and the unfailing support of
many TACs, who continued to provide follow-up time through several years
in the face of poor weather and a lack of interim publications.

\end{document}